\documentclass[acmsmall, screen]{acmart}
\AtBeginDocument{%
  \providecommand\BibTeX{{%
    \normalfont B\kern-0.5em{\scshape i\kern-0.25em b}\kern-0.8em\TeX}}}

\setcopyright{acmcopyright}
\copyrightyear{2026}
\acmYear{2026}

\graphicspath{{figs/}}

\newcommand{\tool}{\textsc{V2XGen}}
\usepackage{graphicx}
\usepackage{subfigure}
\usepackage{booktabs}
\usepackage{bigstrut,multirow,rotating,booktabs}
\usepackage{amsmath}
\usepackage{setspace} 
\usepackage{ulem}
\usepackage{algorithm}
\usepackage{algorithmicx}
\usepackage{algpseudocode}
\usepackage{amsmath}
\floatname{algorithm}{Algorithm}
\usepackage{bbm}
\usepackage{makecell}
\usepackage{array}
\usepackage{enumitem}
\usepackage{lineno}
\usepackage{wrapfig}

\newcommand{\pmm}[1]{\,$\scriptscriptstyle\pm#1$}
\acmJournal{TOSEM}
\acmVolume{0}
\acmNumber{0}
\acmArticle{1}
\acmMonth{0}

\begin{document}

\title{Generating Realistic Test Scenes for V2X Communication Systems}


\author{An Guo}
\email{an.guo@polyu.edu.hk}
\affiliation{
  \institution{The Hong Kong Polytechnic University}
  \city{Hong Kong SAR}
\country{China}
}
\orcid{0009-0005-8661-6133}

\author{Xinyu Gao}
\email{xygao@swjtu.edu.cn}
\affiliation{
  \institution{Southwest Jiaotong University}
  \city{Chengdu}
  \country{China}
}
\orcid{0009-0004-7135-1833}

\author{Chunrong Fang}
\authornote{Chunrong Fang and Zhenyu Chen are the corresponding authors.}
\email{fangchunrong@nju.edu.cn}
\affiliation{
  \institution{State Key Laboratory for Novel Software Technology, Nanjing University}
  \city{Nanjing}
\country{China}
}
\orcid{0000-0002-9930-7111}

\author{Haoxiang Tian}
\email{haoxiang.tian@ntu.edu.sg}
\affiliation{
  \institution{Nanyang Technological University}
  \city{Singapore}
\country{Singapore}
}
\orcid{0000-0001-9132-9319}

\author{Weisong Sun}
\email{weisong.sun@ntu.edu.sg}
\affiliation{
  \institution{Nanyang Technological University}
  \city{Singapore}
\country{Singapore}
}
\orcid{0000-0001-9236-8264}

\author{Yanzhou Mu}
\email{602022320006@smail.nju.edu.cn}
\affiliation{
  \institution{State Key Laboratory for Novel Software Technology, Nanjing University}
  \city{Nanjing}
\country{China}
}
\orcid{0000-0003-1816-2246}

\author{Shuncheng Tang}
\email{sctang@whut.edu.cn}
\orcid{0000-0002-3019-2598}
\affiliation{
\institution{Wuhan University of Technology}
  \city{Wuhan}
  \country{China}
}


\author{Lei Ma}
\email{ma.lei@acm.org}
\affiliation{
  \institution{The University of Tokyo}
  \city{Tokyo}
\country{Japan}
}
\affiliation{
  \institution{University of Alberta}
  \city{Edmonton, Alberta}
\country{Canada}
}
\orcid{0000-0002-8621-2420}

\author{Xiapu Luo}
\email{csxluo@comp.polyu.edu.hk}
\affiliation{
  \institution{The Hong Kong Polytechnic University}
  \city{Hong Kong SAR}
\country{China}
}
\orcid{0000-0002-9082-3208}

\author{Zhenyu Chen}
\authornotemark[1]
\email{zychen@nju.edu.cn}
\affiliation{
  \institution{Shenzhen Loop Area Institute}
  \city{Shenzhen}
\country{China}
}
\orcid{0000-0002-9592-7022}

\renewcommand{\shortauthors}{An Guo, et al.}

\begin{abstract}

Accurately perceiving complex driving environments is essential for ensuring the safe operation of autonomous vehicles. With the tremendous progress in deep learning and communication technologies, cooperative perception with Vehicle-to-Everything (V2X) technologies has emerged as a solution to overcome the limitations of single-agent perception systems in perceiving distant objects and occlusions. Despite the considerable advancements, V2X cooperative perception systems require thorough testing and continuous enhancement of system performance. Given that V2X driving scenes entail intricate communications with multiple vehicles across various geographic locations, creating V2X test scenes for these systems poses a significant challenge. Moreover, current testing methodologies rely on manual data collection and labeling, which are both time-consuming and costly.

In this paper, we design and implement \tool, an automated testing generation tool for V2X cooperative perception systems. \tool~utilizes a high-fidelity approach to generate realistic cooperative object instances and strategically place them within the background data in crucial positions. Furthermore, \tool~adopts a fitness-guided V2X scene generation strategy for the transformed scene generation process and improves testing efficiency. We conduct experiments on \tool~using multiple cooperative perception systems with different fusion schemes to assess its performance. The experimental results demonstrate that \tool~is capable of generating realistic test scenes and effectively detecting erroneous behaviors in different V2X-oriented driving conditions. Furthermore, the results validate that retraining systems under test with the generated scenes can enhance average detection precision while reducing occlusion and long-range perception errors.

\end{abstract}

\ccsdesc[500]{Software and its engineering~Software testing and debugging}

\keywords{Software testing, automated test generation, V2X cooperative driving}


\maketitle

\section{Introduction} 

Autonomous vehicles are increasingly becoming an integral part of daily life~\cite{DBLP:journals/eswa/ZhaoZDWZZCNLB24}. As a prime example of safety-critical intelligent software, autonomous driving systems (ADSs) depend on perception components to interpret information about the surrounding environment and relay perception results to downstream decision modules, thereby ensuring the smooth operation of the system. Recent advancements in deep learning and sensor technology have significantly enhanced the performance of modern perception systems~\cite{li2020lidar,wen2022deep}. Despite these remarkable developments, single-agent perception systems often produce inaccurate or incomplete results due to inherent single-view constraints, such as occlusion and sparse sensor observations at a distance. These limitations can lead to incorrect system behavior or even severe accidents~\cite{crash,DBLP:journals/tosem/KochantharaSFC24,DBLP:conf/kbse/00020TFSSGL0024}.

Vehicle-to-everything (V2X) technology has garnered significant attention in recent years due to its potential to overcome the limitations of single-agent perception in detecting occlusions and long-range obstacles~\cite{cui2022cooperative,ma2024macp,DBLP:journals/corr/abs-2401-01544,DBLP:conf/kbse/GuoZTGPTMWFC25}.
V2X communication systems enable autonomous driving systems to leverage cooperative perception strategies to integrate additional perception information from diverse viewpoints provided by cooperative vehicles~(CVs) and traffic infrastructure, thereby enhancing comprehensive environmental awareness. Numerous leading manufacturers and organizations, including Audi, Tesla, and Baidu, are actively developing V2X platforms to enable autonomous driving systems to accurately perceive their surroundings in challenging traffic scenes~\cite{apollo-v2x, tesla-v2x}.

To ensure continuous improvement of the V2X system, developers must build two typical types of V2X application scenes (i.e., occlusion and long-range perception in the ego vehicle's viewpoint) to iteratively test and optimize V2X systems' corresponding basic functionality~\cite{DBLP:journals/corr/abs-2401-01544,han2023collaborative,DBLP:journals/corr/abs-2310-03525}. However, in practice, building test scenes for V2X systems is a labour-intensive and time-consuming task. It requires practitioners to deploy specific V2X-related infrastructure during real-world road testing to collect data from various traffic scenes~\cite{xu2023v2v4real,DBLP:conf/cvpr/YuLSHYSGLHYN22}. In addition, they must manually annotate the data collected from cooperative vehicles (e.g., images, point clouds, etc.) and thoroughly examine the visualization results to ensure that the scenes are annotated correctly~\cite{DBLP:conf/issta/GuoG00LGSF24,han2023collaborative}. Despite these efforts, the collected scenes struggle to support testing the V2X system adequately due to the large input space, which encompasses various sensor data from individual agents. Thus, there is an urgent need for a more efficient approach to building test scenes.

Software engineering researchers have proposed several testing techniques by generating test scenes to assess the potential risks of perception systems in the real world~\cite{DBLP:conf/icse/LinLZZF22,DBLP:conf/issta/GuoF022,christian2023generating, DBLP:conf/issta/GuoG00LGSF24,DBLP:journals/tse/GuoSGFWTWMC26}. However, these methods~\cite{DBLP:conf/icse/LinLZZF22,DBLP:conf/issta/GuoF022,christian2023generating,DBLP:journals/tse/GuoSGFWTWMC26} mainly focus on testing single-agent perception systems and are difficult to apply to V2X scenes. Due to their disregard for the involvement of cooperative vehicles, these methods face challenges in generating test scenes related to the basic V2X functionalities that are of interest to developers (e.g., the ego vehicle encounters occlusion challenges while the cooperative vehicle has an unobstructed view). Moreover, testing V2X systems requires that the generated test data for different agents be consistent with their perspectives (e.g. a vehicle should appear differently because of the varying viewing angles of traffic participants). Due to the inherent nature of V2X, which involves multi-vehicle collaboration in heterogeneous and complex environments, simply applying a single-agent test method to each agent individually cannot guarantee this consistency, resulting in unrealistic and implausible generated data. Recent work on multi-agent cooperative perception testing~\cite{DBLP:conf/issta/GuoG00LGSF24} primarily focuses on evaluating the robustness of V2X systems under extreme environments and communication conditions (e.g., transmission interference). However, the functional correctness of the system itself, namely its ability to effectively utilize multi-agent information to detect occluded or distant objects, has not been thoroughly evaluated.

To address these challenges, we propose \tool, an automated testing method that generates realistic and V2X-interested scenes for V2X cooperative perception systems. \tool~incorporates five realistic transformation operators grounded in metamorphic testing theory, thereby minimizing the cost of manual data collection and labeling. Each operator is carefully designed, taking into account the operational mechanisms of V2X systems in the physical world, to ensure that the transformed scenes are semantically plausible. By flexibly utilizing and combining these operators, \tool~can manipulate entities within the original data or insert new entities to generate new realistic scenes. 
Moreover, \tool~employs a fitness-guided approach to guide entity manipulation, facilitating the generation of more V2X-interested and challenging test cases for the systems under test.
Finally, \tool~leverages metamorphic relations between the synthesized data and the seed data to automatically identify erroneous perception results.

To evaluate the effectiveness and efficiency of \tool~, we conduct experiments on six popular cooperative perception systems with distinct information fusion schemes to assess the efficacy of \tool. Qualitative and quantitative experiments demonstrate that \tool~can generate realistic test data that meet test input specifications for cooperative perception systems. Our subsequent experimental results reveal that \tool, when using the fitness-guided V2X scene generation method, produces more failed tests on cooperative perception systems compared to CooTest~\cite{DBLP:conf/issta/GuoG00LGSF24} and \tool\_N (w/o fitness-guided strategy) baselines. We further retrain the selected cooperative perception systems with test data generated by \tool, finding that the system performance (measured by average precision) can be improved.

The main contributions of this paper are summarized as follows:

\begin{itemize}

\item \textbf{Method.} We propose an automated testing approach for V2X cooperative perception systems based on fitness-guided metamorphic testing. Specifically, we leverage several high-fidelity transformation operators to manipulate entities in the test seed to generate realistic scenes. 

\item \textbf{Tool.} We implement the above method into a tool \tool. \tool~is an automated scene-generation tool for testing cooperative perception systems. We have released \tool~and the cooperative scenes generated by \tool~at
\url{https://sites.google.com/view/v2xgen/}. 

\item \textbf{Evaluation.} We conduct extensive experiments to evaluate the performance of \tool~across six cooperative perception systems. The results demonstrate that \tool~can generate realistic test scenes and efficiently detect erroneous system behaviors. Furthermore, retraining the cooperative perception system using the generated test scenes significantly enhances its performance.

\end{itemize}

\section{Background}

\begin{figure}[htbp]
	\centering
    \includegraphics[width=0.95\linewidth]{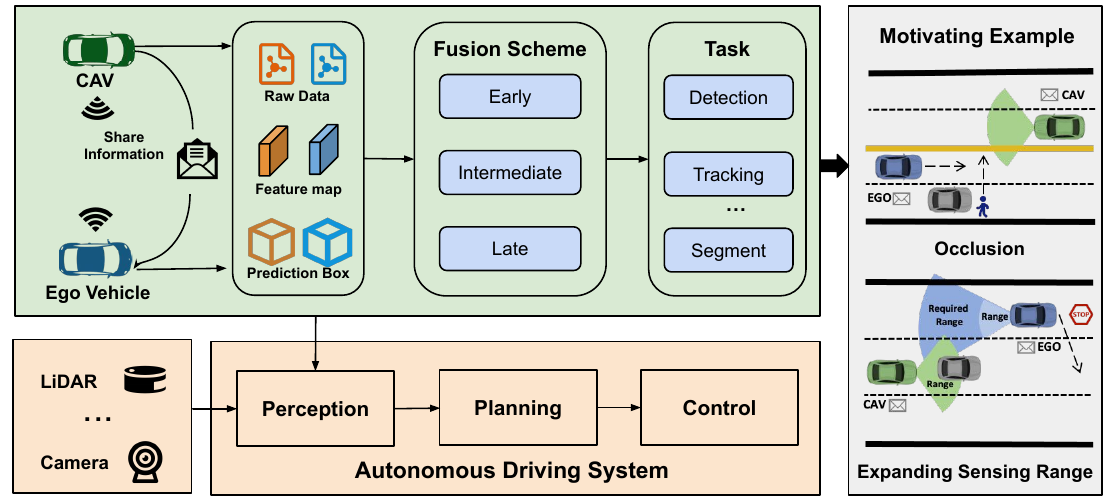}%
	\caption{The architecture of cooperative perception systems in autonomous driving. }
	\label{fig1}
    \vspace{-5pt}
\end{figure}

\subsection{V2X Cooperative Perception System}

Autonomous driving systems consist of sequentially operating subsystems. Figure~\ref{fig1} illustrates the architecture of an autonomous driving software system based on contemporary designs. The perception module is responsible for processing environmental information collected by various sensors to facilitate functionalities such as object detection and tracking. Upon receiving the processed perception data, the planning module estimates future trajectories for the detected obstacles. Finally, the control module calculates and executes the planned trajectory. As the foundation of autonomous driving, perception is essential for environmental understanding.

However, the inherent limitations of camera and LiDAR devices, particularly regarding occlusions and long-distance perception, pose significant challenges for autonomous driving systems designed for individual vehicles~\cite{xu2022opv2v}. These challenges can result in severe consequences in complex traffic scenes. In contrast, cooperative perception enhances multi-vehicle detection by leveraging V2X communication, enabling vehicles and infrastructure to share sensing data. This collaboration provides multiple perspectives on obstacles, compensating for individual limitations \cite{hobert2015enhancements, jung2020v2x}. The integrated data is subsequently utilized by the perception system as input to generate more accurate environmental predictions.

V2X cooperative perception is classified into early fusion, late fusion, and intermediate fusion \cite{han2023collaborative}. Early fusion shares raw data with nearby agents, allowing the ego vehicle to aggregate and predict objects \cite{2019-Cooper}. Late fusion transmits detection outputs and integrates proposals \cite{rawashdeh2018collaborative}. Intermediate fusion balances bandwidth and accuracy by encoding raw data into vector representations before sharing \cite{2019-F-cooper, wang2020v2vnet}.

\begin{figure}[htbp]\small
	\centering
    \includegraphics[width=0.95\linewidth]{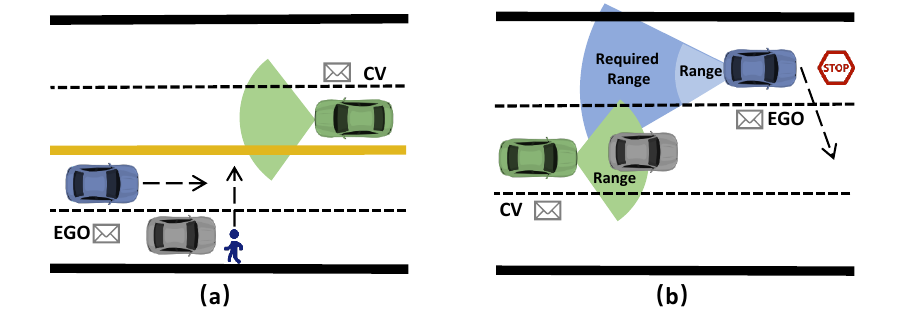}%
	\caption{Motivating examples illustrating the core functionalities of V2X cooperative perception systems.}
	\label{fig:mot}
    \vspace{-5pt}
\end{figure}

\begin{figure*}[htbp]
	\centering
    \includegraphics[width=\linewidth, height=0.35\linewidth]{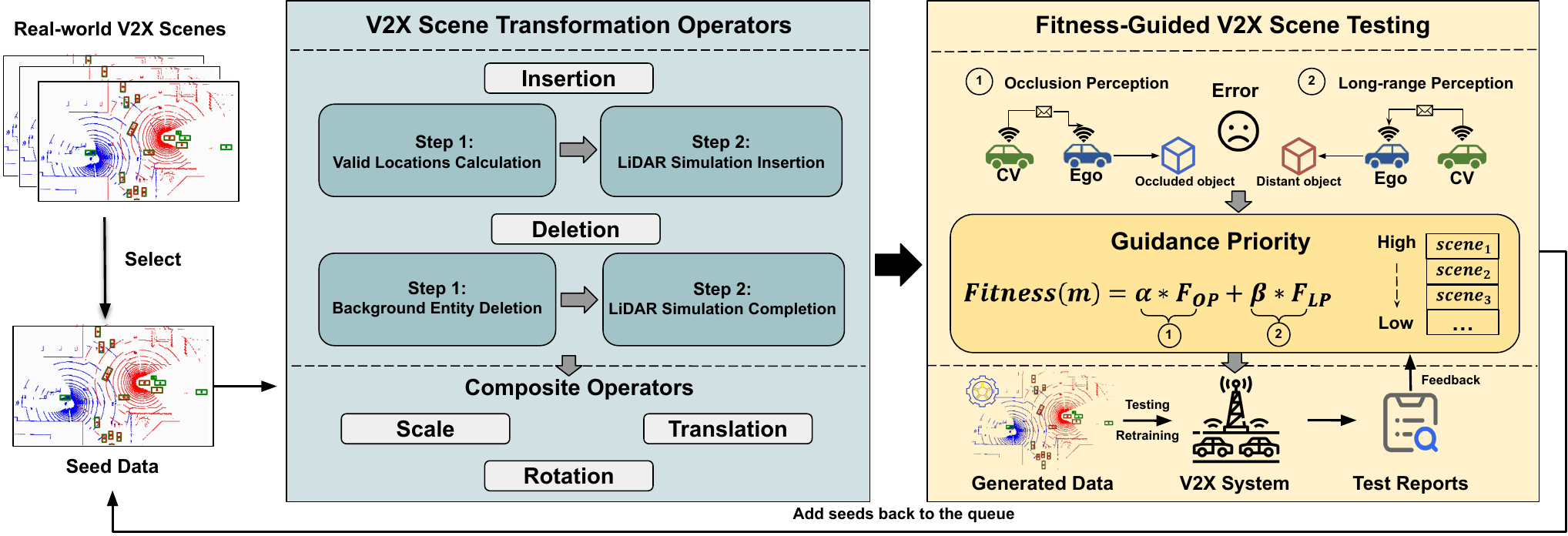}%
	\caption{ The workflow of \tool.}
	\label{fig.v2xframework}
    \vspace{-5pt}
\end{figure*}

\textbf{Motivating Examples.}
The examples in Figure~\ref{fig:mot} highlight the importance of V2X collaborative perception. In Figure~\ref{fig:mot} (a), a green cooperative vehicle in the opposite lane helps the blue ego vehicle detect an occluded pedestrian. 
In Figure~\ref{fig:mot} (b), the blue ego vehicle prepares to change lanes.
The gray vehicle, positioned farther from the ego vehicle, is within the blue ego vehicle’s blind spot and requires assistance from the green cooperative vehicle to enhance its perception range.

\subsection{The Preliminaries of Cooperative 3D Object Detection}
\label{sec2.2}

Cooperative 3D object detection is a fundamental perception task for V2X communication systems, enabling ADS to fully comprehend external environments. In cooperative 3D object detection, the cooperative perception system $CP$ takes a multi-view scene data $s$ as input, which consists of an ego vehicle’s point cloud $P_{ego}$ and point clouds $P_{coo}=\{P_{cv_{i}} \}_{i=1}^n$ from the observation perspectives of $n$ cooperative vehicles $cv_{i}$. The system then outputs a bounding box for each detected object based on the relative position of the ego vehicle, providing its 3D location $loc = [x, y, z]$, dimensions $l$ (length), $w$ (width), $h$ (height), and orientation through a heading angle $yaw$. 
The accuracy of cooperative 3D object detection is typically measured by Intersection over Union (IoU)~\cite{padilla2020survey}. The IoU computation is defined as $I O U=$ area $\left(B_{p} \cap B_{gt}\right) /$ area $\left(B_p \cup B_{gt}\right)$. This metric quantifies the overlap area between a ground-truth 3D bounding box, $B_{gt}$, and a predicted 3D bounding box in bird's-eye view, $B_{p}$, relative to their union~\cite{xu2023v2v4real}. An object is considered successfully detected if the IoU exceeds a specified threshold, $\gamma$. In this study, we set the IoU threshold, $\gamma$, to 0.5, in alignment with prior research~\cite{DBLP:conf/issta/GuoG00LGSF24}.

\section{Approach}

In this section, we present the design and implementation of \tool, a tool devised for automated test generation of the V2X cooperative perception module within ADS. In Figure~\ref{fig.v2xframework}, the developer initiates the process by selecting seeds from real-world V2X scenes. 
Subsequently, \tool~then applies the metamorphic relations specifically defined for cooperative perception systems. 
It is equipped with two basic transformation operators (i.e., insertion and deletion) and three composite transformation operators (i.e., scale, translation, and rotation), which are combinations of the basic operators. The design of these transformation operators in \tool~considers the operational mechanism of the V2X cooperative perception system in the physical environment, allowing for flexible combinations to generate realistic, practical application scenes. Through the use of a fitness metric, \tool~efficiently selects test scenes likely to reveal potential functional errors such as occlusion and long-range perception errors in the systems. The transformed test scenes are then generated to assess the performance of the cooperative perception system in autonomous driving. Finally, \tool~generates test reports containing average precision~(AP), occlusion perception errors, and long-range perception errors based on the system's predicted bounding boxes. Moreover, \tool~can enhance the system's cooperative perception performance through retraining with generated transformed scenes.

\subsection{Definitions}
\label{Definitions}

To efficiently construct challenging test scenes involving occluded and long-range objects, as well as to evaluate the performance of the V2X cooperative perception system in two types of typical application scenes, we define two unique V2X cooperative object detection errors: \textit{\textbf{Occlusion Perception Error (OE)}} and \textit{\textbf{Long-range Perception Error (LE)}}. 
Specifically, occlusion perception error refers to the inability of the cooperative vehicle to assist the ego vehicle in identifying occluded objects. Given an object $o \in \mathbb{O}$ to be detected, a ground-truth bounding box $B_{gt} \in \mathbb{GT}$ and a predicted bounding box $B_p \in \mathbb{D}  \mathbb{T}$ of the cooperative perception system, occlusion perception error can be formalized as:
\begin{equation}
\exists B_{gt} \in \mathbb{G T}, Occ_{ego}(o) >0 \wedge \forall B_p \in \mathbb{D T}, 0\leq IoU\left(B_{gt}, B_p\right) \leq \gamma
\end{equation}

\noindent where $Occ_{ego}(o)$ is the line-of-sight occlusion rate of the ego vehicle detecting object $o$. Occlusion perception error typically occurs when the object $o$ intended for detection is occluded by other obstacles within the ego vehicle's field of view. Long-range perception error refers to the cooperative vehicle's inability to assist the ego vehicle in identifying objects at long distances, which can be formalized as:
\begin{equation}
\label{eq:disfault}
\exists B_{gt} \in \mathbb{G T}, Dis^{ego}(o) >k \wedge \forall B_p \in \mathbb{D T}, 0\leq IoU\left(B_{gt}, B_p\right) \leq \gamma
\end{equation}
\noindent where $Dis^{ego}(o)$ represents the distance between the ego vehicle and the detected object $o$, and $k$ is the distance parameter for long distances in the cooperative perception dataset.


\subsection{V2X Scene Transformation Operators}\label{section:V2X Simulation}

One significant threat to generate realistic test scenes for V2X systems is the requirement to synthesize perspective-consistent scenes for each agent in V2X systems. This means objects in each scene should appear differently depending on the viewpoint of the vehicle.
For instance, as depicted in Figure~\ref{fig.insert} (b), from the ego vehicle's perspective, the yellow car is partially obscured by the grey car, while the green cooperative vehicle has an unobstructed view. 
To address this, we establish perspective transformation between multi-agents and further design five transformation operators for V2X scenes, including two basic operators (entity insertion and deletion) and three composite operators (scale, rotation, and translation) built upon basic operators. Each operator manipulates entities based on multi-agent perspective transformation and real-world physics laws to generate perspective-consistent and realistic test scenes.

\subsubsection{Multi-agent Perspective Transformation}
\label{sec: trans}

The V2X communication system needs to determine the positional transformation matrix between multiple agents to ensure that the perception information measured in different local coordinate systems can be converted to each other. Therefore, we uniformly establish a world coordinate system to represent this transformation.
The transformation from the world coordinates to the vehicle $V_x$ ($x\in \{ego, cv_{i}\}| i \in 1,2,\ldots,n \}$) can be expressed as:
\begin{equation}
\label{eq:trans}
p_{\mathrm{V_x}}^t=T_{\mathrm{world}\rightarrow {V_x}}^t  \cdot p_{\mathrm{{world}}}
\vspace{-3pt}
\end{equation}


\noindent where $p_{\mathrm{{world}}}$ is a location $[\omega_{x_0}, \omega_{y_0}, \omega_{z_0}, 1]^{\mathrm{T}}$ of world coordinate system, $p_{\mathrm{V}_x}^{t}$ denotes the corresponding position in the coordinate system of the participating vehicle $V_x$ at time $t$, and $T_{\mathrm{world}\rightarrow {V_x}}^t \in \mathbb{R}^{3 \times 4}$ is a coordinate transformation matrix from world coordinate system to the local coordinate system of participant $V_x$ at the time $t$. The coordinate transformation matrix $T_{\mathrm{world}\rightarrow {V_x}}^t$ can be represented as:
\begin{equation}
\label{eq:matrix}
T_{\mathrm{world}\rightarrow V_x}^t
=
\begin{bmatrix}
r_{11}^t & r_{12}^t & r_{13}^t & t_x^t \\
r_{21}^t & r_{22}^t & r_{23}^t & t_y^t \\
r_{31}^t & r_{32}^t & r_{33}^t & t_z^t
\end{bmatrix}
\in \mathbb{R}^{3 \times 4}
\end{equation}
where the first three columns constitute a rotation matrix, and the final column represents a translation vector. These values are determined with respect to the local coordinate system established by the data collected by the participating vehicle $V_x$ at time $t$. Based on Equation~\ref{eq:trans}, we can apply the transformation matrix to convert the world coordinates to those of any participating vehicle's perspective. Conversely, the participating vehicle can use the corresponding inverse matrix to transform back to the world coordinate system. Based on this transformation mechanism, the operator we designed strictly ensures perspective consistency. However, in real-world V2X systems, the relative positioning of participating vehicles is estimated, and this process may introduce errors.

\subsubsection{Insertion Operator}
The insertion operator~(IS) is designed to generate plausible new point cloud scenes $s'$ by adding entities into the original multi-view scene $s = \{P_{ego}, P_{cv_{i}}| i \in 1,2,\ldots,n \}$.
Given an entity instance $e_{ins}$ selected from the entity assets $\mathbb{E}$, \tool~first calculates valid locations and orientations of the entity to be inserted for each view. Then \tool~leverages a set of virtual LiDARs to render the entity instance as object-level point clouds and integrate them into the scene at the corresponding viewpoints correctly by processing their interaction with background objects. We detail each step of the insertion operator below.

\noindent\textbf{Step 1. Valid Locations Calculation.} Inserting entities at appropriate locations is crucial for synthesizing realistic test scenes. The location of each inserted entity must be reasonable (e.g., a vehicle should be correctly placed on the road) and adhere to the requirement of the transformation relationship outlined in Section~\ref{sec: trans} across multiple views. 
To compute feasible candidate insertion points, \tool~first employs a high-efficiency semantic segmentation model~\cite{DBLP:conf/isvc/CortinhalTA20} to separate the road point cloud from the ego-perspective point cloud $P_{ego}$ and each cooperative point cloud $P_{cv_i}$, respectively. Then, \tool~computes the insertion point in the world coordinate system and filters out points that are not on the road surface in any of the views based on Equation~\ref{eq:trans} to ensure that all candidate locations are valid from each viewpoint. Moreover, given an entity $e_{ins}$ with specific orientation and dimensions, inserting the entity at a candidate location may result in a collision with the objects in the background scene $s$. To address this, \tool~calculates the 3D bounding box $B_e$ for the entity instance to be inserted and checks if the 3D box $B_e$ contains any other objects in the background scene for each view. If the $IoU$ between $B_e$ and all the background bounding boxes is 0, \tool~considers this selected location to be non-colliding. Besides, to guarantee that at least one cooperative vehicle can detect the inserted entity, \tool~computes occlusion rates of the entity from multiple views and ensures that the occlusion rate from at least one cooperative vehicle’s perspective remains below 90\%.

\begin{figure}[htbp]
	\centering
    \includegraphics[width=\linewidth, height=0.35\linewidth]{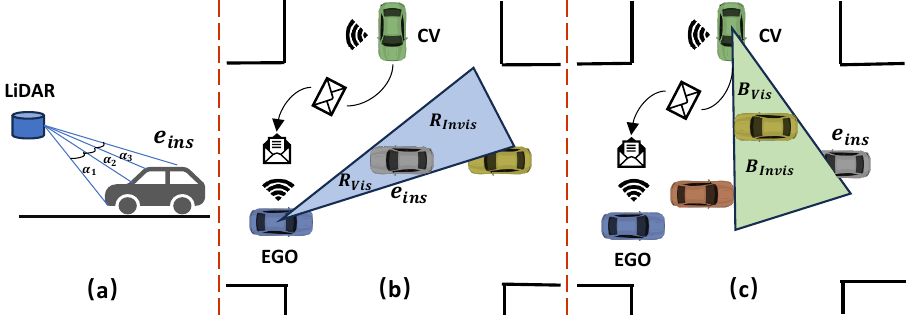}%
        \vspace{-5pt}
	\caption{ Subfigure (a) (y-axis view) illustrates the virtual LiDAR equipped with a V2X cooperative sensing system emitting a laser to hit the inserted entity $e_{ins}$. Subfigure (b) (top view) shows the calculation of the region
    occluded by the newly inserted entity $e_{ins}$  (within region $R_{Invis}$). Subfigure (c) (top view) illustrates the calculation of the region of the inserted entity $e_{ins}$ occluded by the existing background object (within region $B_{Invis}$).}
	\label{fig.insert}
    \vspace{-10pt}
\end{figure}

\noindent\textbf{Step 2. LiDAR Simulation Insertion.} 
Given a candidate location $loc_{ins} = \{loc_{ego},loc_{cv_{i}}|i \in 1,2,\ldots, n \}$ for insertion, where $loc_{ins}$ denotes the set of relative coordinates of the inserted entity with respect to the local coordinate systems of vehicle $V_x$ ($x\in \{ego, cv_{i}\}| i \in 1,2,\ldots,n \}$), this step aims to render the entity instances into original scene $s = \{P_{ego}, P_{cv_{i}}| i \in 1,2,\ldots, n \}$ respectively to generate new perspective-consistency scene $s'$. The realism of the synthesized scene depends on two key factors: ensuring the inserted entities maintain perspective consistency across multiple views, and ensuring these entities conform to the physics laws in each individual view. To address this, we leverage a high-fidelity approach to simulate the operational mechanism of cooperative perception LiDAR in the physical world. 
Specifically, we create a multi-view virtual LiDAR sensor set for the V2X scene generation. Each participant in the system is equipped with a configurable virtual LiDAR, initialized based on the real-world LiDAR configuration in the V2X system (e.g., the field of view, the number of beams, etc).
The virtual LiDAR utilizes a ray-casting algorithm to simulate the real laser beam emission, thereby rendering the inserted entity $e_{ins}$ as an object-level point cloud consistent with the viewpoint.  Figure~\ref{fig.insert}(a) illustrates the ray-casting process of a nonlinear LiDAR in a V2X system. Moreover, V2XGen handles interaction with background objects in the scene at each viewpoint. For each viewpoint, V2XGen performs two operations. First, it removes background point clouds occluded by the inserted entity, as illustrated in Figure~\ref{fig.insert}(b). Second, it removes the occluded parts of the inserted entity based on geometric relationships, as illustrated in Figure~\ref{fig.insert}(c). Figure~\ref{fig.insert}(b) and Figure~\ref{fig.insert}(c) illustrate the processing procedures from the ego viewpoint and the cooperative vehicle viewpoint, respectively. Figure~\ref{fig.insert}(b) illustrates how V2XGen removes the background point cloud occluded by the gray inserted entity $e_{ins}$. After insertion, LiDAR beams emitted from the ego vehicle that would otherwise reach yellow background objects become blocked by $e_{ins}$. Therefore, the point cloud behind $e_{ins}$ within the invisible region $R_{Invis}$ must be removed to maintain physical plausibility. Figure~\ref{fig.insert}(c) illustrates the opposite case, where the gray inserted entity $e_{ins}$ is partially occluded by an existing yellow background object. To preserve insertion realism, V2XGen removes the corresponding occluded parts of $e_{ins}$ within region $B_{Invis}$ from the rendered point cloud.

\subsubsection{Deletion Operator}
The deletion operator~(DL) is designed to generate a realistic new scene by removing entities in multiple views and reconstructing the previously occluded areas. Our core idea is that deletion may alter the occlusion relationships among objects, thereby affecting the reasoning confidence of the cooperative perception system. Given the entity instances selected for deletion from the cooperative perception data recorded in the real world, \tool~first deletes the selected background entity by filtering the points within the object's ground truth box. Then, \tool~leverages the high-fidelity virtual LiDAR simulation to fill the occlusion area of the selected entity. We detail each step of the deletion operator below.

\begin{figure}[htbp]
	\centering
    \includegraphics[width=\linewidth, height=0.36\linewidth]{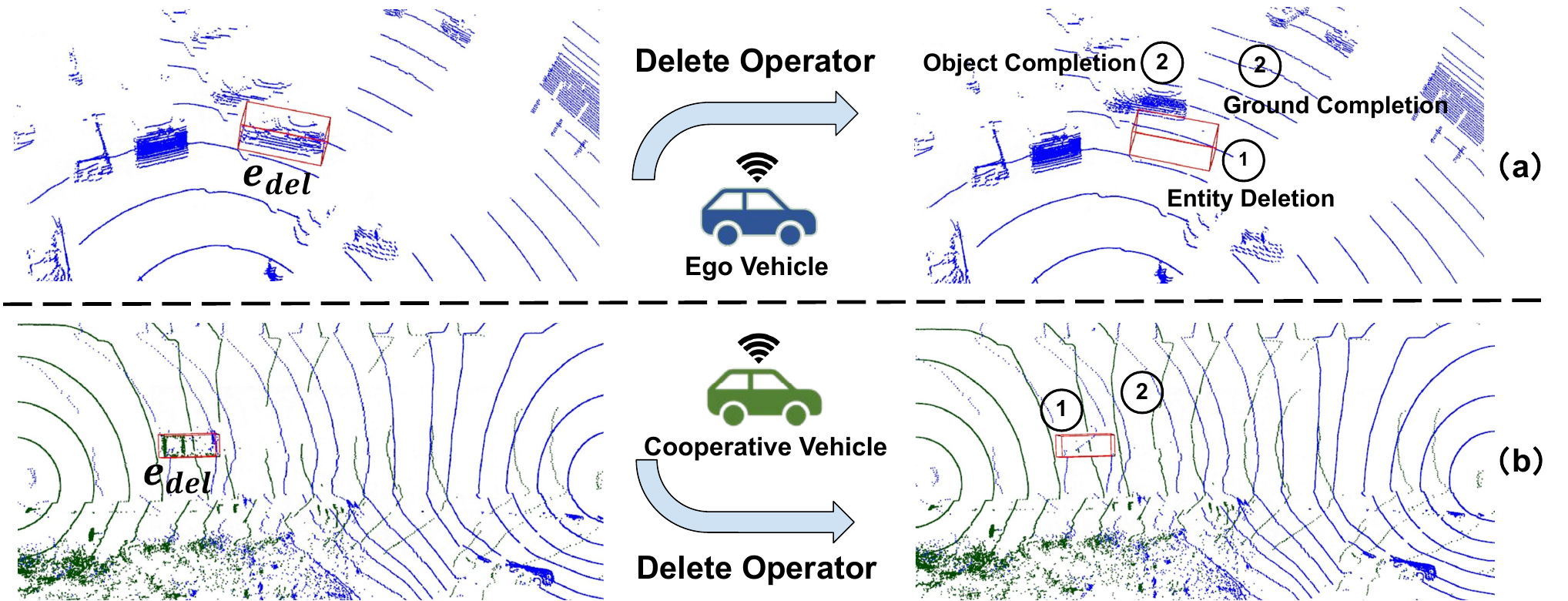}%
	\caption{ Subfigures (a) and (b) illustrate the visualization effects of ego vehicle data and aggregated LiDAR data, respectively, generated by \tool~after the application of the deletion operator. The entities selected for deletion by \tool~are marked in red. }
	\label{fig.delete}
    \vspace{-10pt}
\end{figure}

\noindent\textbf{Step 1. Background Entity Deletion.} \tool~first randomly selects a background entity instance $e_{del}$ for deletion from the cooperative perception scene and determines candidate location $loc_{del} = \{loc_{ego}, loc_{cv_{i}}|i \in 1,2,\ldots, n \}$ for deletion based on Equation~\ref{eq:trans}. Here, the background entity refers to an existing object instance in the original scene that serves as the deletion target of the deletion operator. $loc_{del}$ denotes the set of relative coordinates of the target background entity to be deleted with respect to the local coordinate systems of each vehicle $V_x$ ($x\in \{ego, cv_{i}\}| i \in 1,2,\ldots,n \}$). The candidate set for deletion consists of background entities belonging to non-connected vehicles. Then, for each participant's viewpoint, \tool~calculates the entity's 3D bounding box $B_e$ based on ground truth and deletes all point clouds contained within $B_e$ to remove the background entity.

\noindent\textbf{Step 2. LiDAR Simulation Completion.} After deleting background entities' point clouds, \tool~leverages a high-fidelity occlusion-completion strategy based on virtual sensor simulation to reconstruct the previously occluded background ground and objects. We leverage the multi-view virtual LiDAR sensor set to infer the invisible occlusion region of the selected entity instance $e_{del}$ and sequentially complement the ground and objects in the invisible area. To complete the ground, we meshify the road point cloud to reconstruct the road surface and leverage the ray-casting algorithm to complete the obscured ground with realistic scan lines. The accuracy of road surface reconstruction directly affects the quality of ground completion, with higher reconstruction accuracy leading to improved completion results. In order to complete the background objects, we compute the intersection of the occluded region with the bounding boxes of all background objects to identify those that were occluded by the deleted entity $e_{del}$. We then reinsert entities of the same size as the originals at the locations of each occluded object and leverage the multi-view virtual LiDAR sensor set to reconstruct the occluded object under each view. Higher-quality 3D models of the reinserted entities lead to more accurate completion of the background objects. Figure~\ref{fig.delete} presents the visualization of ego-vehicle data and aggregated LiDAR data before and after entity deletion. The visualization results indicate that our approach is capable of performing entity deletion and LiDAR simulation completion.

\subsubsection{Composite Operators}

In the scene of cooperative perception, the shapes of objects within a single class can vary significantly (e.g., flat sports cars and round vintage cars~\cite{DBLP:conf/cvpr/WangCYLHCWC20}). Such long-tail data have the potential to be misclassified. Additionally, dynamic changes in the heading directions and locations of objects can adversely affect the positioning accuracy of cooperative perception systems~\cite{li2023common, DBLP:conf/icra/MillerRCKL20}. Therefore, we utilize scale, rotation, and translation operators to simulate objects with various shapes, orientations, and positions in the real world. We regard these operators as a combination of insertion and deletion operators to ensure the realism of the generated scenes. The combined operators inherently conform to the laws of physics and maintain perspective consistency since each basic operator in the composition is rigorously designed to ensure these properties. We provide the details of each operator below.

\noindent\textbf{Scale Operator.}
The scale operator~(SC) changes the shape of the object. The operator first randomly selects the entity to be scaled from the multi-view cooperative perception scene $s$. For each participant’s viewpoint, it deletes an entity with length $x$, width $y$, and height $z$ and then inserts a new entity with a length $s_x*x$, width $s_y*y$, and height $s_z*z$ respectively, where $s_x$, $s_y$, and $s_z$ are the scale ratios. To ensure realism, this operator sets conservative scale parameters ($s_x \in [0.9,1.1], s_y \in [0.9,1.1], s_z \in [0.9,1.1]$) and automatically calculates the position of the entity to ensure that it is placed on the ground.

\noindent\textbf{Rotation Operator.}
The rotation operator~(RO) changes the direction of the object. The operator first randomly selects the entity to be rotated from the multi-view cooperative perception scene $s$. For each participant’s viewpoint, it deletes the entity with the original orientation of $r_z$ and inserts the entity with the orientation of $r_z + rot$. $rot$ is the rotation angle around the upright $z$ axis. To ensure realism, the operator sets conservative rotation parameters ($rot \in [-30^\circ,-5^\circ]\cup[5^\circ,30^\circ]$) to ensure that the original movement direction of the entity is not changed.

\noindent\textbf{Translation Operator.}
The translation operator~(TR) changes the position of the object. The operator first randomly selects the entity to be translated from the multi-view cooperative perception scene $s$. For each participant’s viewpoint, it deletes the entity with the original position $(x, y, z)$ and inserts the entity with the new position $(x + t_x, y + t_y, z')$. $t_x$ and $t_y$ are the translation distances of the object along the $x$ and $y$ axes, respectively. To ensure that an entity is positioned on the ground, the $z'$ coordinate of the translated object is automatically calculated based on its coordinates on the $x$ and $y$ axes following the translation. Since the translation operator utilizes an insertion operation, it guarantees that the entity's translated position resides on the road.

\vspace{-5pt}

\subsection{Fitness-Guided V2X Scene Testing}\label{section:V2X Simulation}

In theory, the number of tests each operator can generate is endless. Therefore, we must impose limitations and guidance to ensure test quality and improve testing efficiency. To enhance the testing efficiency of \tool, we propose a fitness-guided testing process for V2X cooperative perception systems under the interactive cooperation of $n$ vehicles. The core idea of this guidance strategy is to generate scenes with occlusion relationships and long-range objects from the perspective of the ego vehicle while maximally confusing cooperative perception systems. Next, we introduce the fitness-guided strategy in detail.

\noindent\textbf{Fitness Metric.} V2X cooperative perception is designed to enhance perception under challenging conditions, particularly in terms of occlusion handling and long-range object perception (as shown in Figure~\ref{fig:mot}). In realistic cooperative perception scenarios, these two challenges may occur simultaneously, as a scene can contain both occluded and long-range objects. Therefore, an effective fitness metric should jointly evaluate the error-revealing potential of test data with respect to both capabilities during the generation process, while allowing their relative importance to be flexibly adjusted~\cite{feng2020deepgini}. To this end, we design a fitness metric that measures the likelihood of test data revealing perception errors. Our fitness metric comprises two error-revealing ability scores: $F_{OP}$, designed to detect occlusion perception errors, and $F_{LP}$, intended to identify long-range perception errors.

Given a ground-truth bounding box $B_{gt} \in \mathbb{G T}$ and a predicted bounding box $B_p \in \mathbb{D} \mathbb{T}$, the occlusion perception error score can be expressed as follows:
\begin{equation}\small
\begin{aligned}
F_{OP}=\sum_{B_{gt} \in \mathbb{GT}} I_{G T}\left(B_{gt}, B_p\right) * Occ_{ego}(o_{gt}) *\prod_{i=1}^{n} \left(1-Occ_{cv_i}(o_{gt})\right)
\end{aligned}
\end{equation}

\noindent where $I_{G T}(\cdot)$ is an indicator function equal to 1 if and only if the cooperative perception system fails to detect the object $B_{gt}$. $Occ_{ego}(o_{gt})$ and $Occ_{cv_i}(o_{gt})$ represent the maximum line-of-sight occlusion rate for the ego vehicle and the i-th cooperative vehicle, respectively, when detecting the object $o_{gt}$ with the ground truth bounding box $B_{gt}$. To detect obstacle $o_{gt}$ through cooperative perception, a cooperative vehicle with a low occlusion rate should assist the ego vehicle in detecting obstacles that are highly occluded from the ego vehicle’s viewpoint. Otherwise, the cooperative perception system is considered inadequate in handling occluded obstacles. Accordingly, to generate challenging scenes with occlusion relationships, we prioritize retaining test scenes in which the ego vehicle has a limited field of view while all cooperative vehicles maintain high visibility of the obstacle $o_{gt}$. Conversely, we discard scenes in which at least one cooperative vehicle has zero visibility of the obstacle $o_{gt}$ or in which the ego vehicle has full visibility. In the former case, obstacle $o_{gt}$ is completely occluded from the cooperative vehicle’s view, preventing it from assisting the ego vehicle in detection. In the latter case, the ego vehicle already has an unobstructed view of obstacle $o_{gt}$, with no occlusion.

The long-range perception error score is expressed as:
\begin{equation}\small
\begin{aligned}
F_{LP}=\sum_{B_{gt} \in \mathbb{GT}} I_{G T}\left(B_{gt}, B_p\right)
*\frac{\min \left(D i s^{ego}\left(B_{gt}\right), D i s^{ego}_{max} \right)}{D i s^{ego}_{max} }
*\prod_{i=1}^{n}{(1-\frac{\min \left(D i s^{cv_i}\left(B_{gt}\right), D i s^{cv_i}_{max} \right)}{D i s^{cv_i}_{max} })}
\end{aligned}
\end{equation}

\noindent where $dis^{ego}(.)$ calculates the distance between a bounding box $B_{gt}$ and the ego vehicle's LiDAR position, $dis^{cv_i}(.)$ calculates the distance between a bounding box $B_{gt}$ and the i-th cooperative vehicle's LiDAR position, $d i s^{ego}_{max}$ and $d i s^{cv_i}_{max}$ are the maximum recognition distances of the LiDAR equipped by the ego vehicle and the i-th cooperative vehicle respectively. The intuition behind the score is that the cooperative perception system should aid the ego vehicle in detecting distant objects. We aim to investigate scenes in which a nearby cooperative vehicle offers limited assistance to the ego vehicle, potentially resulting in false detections.

Finally, our fitness metric can be expressed as a weighted sum of the two scores:
\begin{equation}
\vspace{-3pt}
Fitness(m)=\alpha * F_{OP}+\beta * F_{LP}
\label{equa5}
\end{equation}
\noindent where $\alpha+\beta=1$. Similar to the traditional software testing approach that aims to increase code coverage, \tool~attempts to generate a test set designed to maximize the fitness score.


    \begin{algorithm}[!htb]\small
        \caption{Fitness-guided V2X Scene Generation}
        \begin{algorithmic}[1] 
            \Require {
            The tested system $CP$, the transformation operator set $\mathbb{A}$, the set of seed scene $\mathbb{S}_{ori}$, the number of generated tests $gen_{num}$, the maximum number of entity manipulations $N$ }
            \Ensure  Generated high-quality test data $\mathbb{TD}$
            \\
            $\mathbb{TD} \leftarrow \emptyset$\;
            \For{$s$ in $\mathbb{S}_{ori}$}
                \State $\mathbb{GT}\leftarrow \mathit{LoadLabel}(s)$
                \State $s^{\prime},\mathbb{GT}^{\prime} \leftarrow s,\mathbb{GT}$
                \For{$i$=1,2\ldots,$N$}
                \State $a=\mathit{RandSelOperator}(\mathbb{A})$
                \State $s^{\prime},\mathbb{GT}^{\prime}=\mathit{Transform}(s^{\prime},\mathbb{GT}^{\prime},a)$
                \EndFor
                \State $\mathbb{DT}^{\prime}=\mathit{GetPred}(CP,s^{\prime})$
                \State $Occ, Dis=\mathit{CalOccandDis}(\mathbb{GT}^{\prime}, s^{\prime})$
                \State $Fitness_{pri}=\mathit{CalFitnessPri}(\mathbb{DT}^{\prime},
                \mathbb{GT}^{\prime},Occ, Dis)$
                \State $Spc=\mathit{CreateTransScene}(s^{\prime},\mathbb{GT}^{'},Fitness_{pri})$
                \If{$len(\mathbb{TD})<gen_{num}$}
                \State $\mathbb{TD}.update(Spc)$
                \State $\mathbb{TD}.\mathit{SortByFitness()}$
                \Else
                    \If{$Spc.Fitness_{pri}>\mathbb{TD}[-1].Fitness_{pri}$}
                    \State $\mathbb{TD}[-1]=Spc$
                    \State $\mathbb{TD}.\mathit{SortByFitness()}$
                    \EndIf
                \EndIf
            \EndFor
            \\
            \textbf{return} $\mathbb{TD}$
        \end{algorithmic}
        \label{alg1}
    \end{algorithm}

\noindent\textbf{Testing Workflow.} 
This fitness-guided strategy enables \tool~to uncover more system defects with the same testing set size, thereby improving testing efficiency. 
Algorithm~\ref{alg1} presents the process of fitness-guided V2X scene generation in detail. The algorithm takes the tested system $CP$, a set of scene transformation operators $\mathbb{A}$ with the corresponding parameters, a seed set $\mathbb{S}_{ori}$, the number of generated tests $gen_{num}$, and the maximum number of object mutations $N$ as input. The algorithm first loads the label of the scene data $s$ to be transformed (line 3). 
Next, it randomly selects a transformation operator from the predefined operator set (line 6). It then randomly selects an object from the scene, renders the object's point cloud using a virtual LiDAR sensor, integrates the rendered point cloud with the background data to synthesize the transformed scene $s^{\prime}$, and generates the corresponding label $\mathbb{GT}^{\prime}$ for $s^{\prime}$ (line 7).
The core implementation process of this algorithm includes first putting the seed $s^{\prime}$ into the cooperative perception system for prediction. Subsequently, it calculates the distance and occlusion relationship of objects (lines 9-10), followed by the computation of the fitness guidance metric value based on the aforementioned relationship and bounding box prediction (lines 11-12). Then, the generated test cases will be added to $\mathbb{TD}$ (lines 13-15). When the generated quantity surpasses $gen_{num}$  (lines 16-21), \tool~retains tests that are more likely to detect occlusion perception errors and long-range perception errors. Note that, to keep the retention logic of the priority queue $\mathbb{TD}$ concise, $Spc$ is treated as a structured object. Specifically, $Spc$ denotes the composite test-case object constructed on line 12, which encapsulates the transformed scene $s^{\prime}$, its ground-truth labels $\mathbb{GT}^{\prime}$, and the associated fitness score $Fitness_{pri}$.

\subsection{Metamorphic Relations}\label{section:MR}

Manual labeling of our generated test cases can be time-consuming and labor-intensive. To alleviate this, we leverage metamorphic relations (MRs)~\cite{chen2018metamorphic, 2020-Metamorphic-Testing} to create test oracles. MRs describe the necessary properties of a target software in terms of inputs and their expected outputs. Violations of MRs often indicate potential errors.

\tool~is designed for cooperative perception systems based on metamorphic testing. Specifically, we denote the cooperative perception system as $C P$, which detects results using the multi-view cooperative perception scene $s \in \mathbb{S}$, including point clouds collected from ego and cooperative vehicles. Given a set of entity instances $\mathbb{E}_{ins}$, an insertion MR can be formalized as follows:
\begin{equation}\small
M R_1: \forall e_{ins} \in \mathbb{E}_{ins} \wedge \forall s \in \mathbb{S}, \zeta\left\{C P \llbracket s \rrbracket \cup G T_{ins}, C P \llbracket \sigma(s, e_{ins}) \rrbracket\right\}
\label{eqa7}
\end{equation}
\noindent where $\sigma$ is the insertion operator for inserting an entity instance $e_{ins}$ into the multi-view background scene $s$, $G T_{ins}$ is the ground truth of the entity $e_{ins}$ (i.e., the estimated bounding box in object detection task) and $\zeta$ is a criterion asserting the equality of $C P$ results. Similar to the insertion MR, given the set of background entity instances $\mathbb{E}_{del}$ from $s \in \mathbb{S}$, the deletion MR used to test $C P$ can be represented as follows:
\begin{equation}\small
M R_2: \forall e_{del} \in \mathbb{E}_{del} \wedge \forall s \in \mathbb{S}, \zeta\left\{C P \llbracket s \rrbracket \setminus G T_{del}, C P \llbracket \tau(s, e_{del}) \rrbracket\right\}
\label{eqa8}
\end{equation}
\noindent where $\tau$ is the deletion operator for deleting an entity instance $e_{del}$ from the multi-view background scene $s$, $G T_{del}$ is the ground truth of the entity $e_{del}$.

Regardless of which vehicle is treated as the ego vehicle, the cooperative perception system should satisfy $M R_1$ and $M R_2$. Otherwise, the cooperative perception system is considered to exhibit erroneous behaviors. $M R_1$ and $M R_2$ are built upon the following two facts: (1) the object insertion and deletion operators should not change the correct prediction of $C P$; (2) the inserted object should be detected correctly, and the deleted object should not be detected. 
Since the remaining composite operators are combinations of insertion and deletion operators, their $M Rs$ can be automatically derived based on $MR_1$ and $MR_2$. However, since transformation operators may introduce noise or unrealistic artifacts, V2XGen ensures semantic validity by filtering out invalid generated scenes. For each entity transformation (e.g., insertion or translation), the modified entity must be successfully detected by the single-agent perception model of either the ego vehicle or any cooperative vehicle. Otherwise, the transformation is deemed semantically invalid, and the generated scene is discarded and the transformation is re-executed. Additionally, asserting equality according to Equation~\ref{eqa7} and Equation~\ref{eqa8} is too strict and can lead to a large number of false positives due to a slight drift in detection results. Therefore, we follow previous work~\cite{DBLP:conf/issta/GuoF022} to leverage soft equality criteria $\zeta$ derived from Average Precision (AP). Given the MR, we can obtain the test oracle information without manual annotation by checking if the $\mathrm{MR}$ is violated.

\section{Experimental Design}

\tool~is designed to systematically test the cooperative driving system, particularly to verify its performance on long-range and occlusion perception scenes. To this end, we empirically explore the following three research questions (RQ):

\begin{itemize}
\item RQ1: How effective is the \tool~at synthesizing realistic cooperative perception data?
\item RQ2: How effectively can \tool~generate tests under the fitness-guided scene generation?
\item RQ3: How effective is \tool~in enhancing cooperative perception systems through retraining?
\end{itemize}

\subsection{Implementation Details}

In all experiments, we set the maximum number of entity manipulations $N$ to 3 in Algorithm~\ref{alg1}. $\alpha$, $\beta$ in Equation~\ref{equa5} are set to 0.5 and 0.5, respectively. Given that the V2V4Real~\cite{xu2023v2v4real} dataset is collected using multiple vehicles outfitted with Velodyne VLP-32 LiDAR sensors, we adopt the Velodyne VLP-32 LiDAR configuration for the construction of our virtual simulator. To match the 64-beam vehicle-mounted LiDAR data in V2X-Real~\cite{DBLP:conf/eccv/XiangZXXGZHJLMJLMHMYZM24}, we adopted a 64-beam LiDAR configuration when constructing the virtual simulator. The distance parameter $k$ in Equation~\ref{eq:disfault} is set to 50 (m) according to the distance division of the V2V4Real dataset. We implement \tool's workflow to conduct the experiments.

\subsection{Datasets}
We conduct experiments using the V2V4Real~\cite{xu2023v2v4real} and V2XReal~\cite{DBLP:conf/eccv/XiangZXXGZHJLMJLMHMYZM24} test datasets under the default configuration, with one agent randomly selected as the ego vehicle. V2V4Real is a real-world dataset for Vehicle-to-Vehicle cooperative perception, comprising sensor data from cooperative vehicles across diverse scenarios. It includes 347 km of highway and 63 km of city driving logs, 20,000 LiDAR frames, 240,000 annotated 3D bounding boxes, and HD maps of all routes. V2X-Real is a large-scale real-world dataset for V2X cooperative perception, comprising synchronized sensor data collected in challenging urban environments. The full dataset contains 33,000 LiDAR frames and more than 1.2 million annotated 3D bounding boxes. For each dataset, we apply our proposed technique to the original test split of the V2V cooperative object detection benchmark to identify errors in the V2X cooperative perception system. Additionally, we use ShapeNet~\cite{chang2015shapenet} to construct entity assets $\mathbb{E}$ for insertion. ShapeNet is a large-scale dataset of over 220,000 annotated 3D object models. For this study, we build an entity asset from the car category in ShapeNet, which consists of 3,483 entities with various styles.

\subsection{V2X Cooperation Perception Systems}

To comprehensively evaluate the performance of our techniques, we utilize six cooperative perception systems with diverse fusion schemes in our experiment. These systems consist of one early fusion system~\cite{xu2023v2v4real}, one late fusion system~\cite{xu2023v2v4real}, and four state-of-the-art intermediate fusion systems (V2VNet~\cite{wang2020v2vnet}, V2X-ViT~\cite{xu2022v2x}, AttFusion~\cite{xu2022opv2v}, F-Cooper~\cite{2019-F-cooper}). All of these perception systems are based on LiDAR and are designed to process multi-view point clouds as input. Due to the page limit, we refer audiences to our supplementary website~\cite{website} for details of each system under test.

\subsection{Evaluation Metric and Baseline Comparison}\label{section:metric}

\noindent\textbf{Metric.} 
The goal of cooperative perception is to collaboratively detect objects across multiple 3D point clouds from the ego and cooperative vehicles. To evaluate cooperative 3D object detection, we first apply the IoU metric (Section~\ref{sec2.2}) to determine successful detections. We then assess performance using Average Precision (AP), a standard metric in V2X cooperative detection research~\cite{xu2023v2v4real,DBLP:conf/issta/GuoG00LGSF24}. For fair comparisons, we adopt the 11 recall positions from the Pascal VOC benchmark~\cite{everingham2010pascal}, defined as follows:
\begin{equation}\small
\left.\mathrm{AP}\right|_{R}=\frac{1}{|R|} \sum_{r \in R} \rho_{\text {interp }}(r)
\vspace{-3pt}
\end{equation}
We use 11 equally spaced recall levels, denoted as $R_{11}=\{0, 0.1, 0.2, \ldots, 1\}$. The interpolation function, $\rho_{\text{interp}}(r)=\max _{r^{\prime}: r^{\prime} \geq r} \rho\left(r^{\prime}\right)$, is employed, where $\rho(r)$ represents the precision at recall $r$. A higher AP reflects better performance of cooperative perception systems.

To evaluate the realism of the cooperative perception data, we utilize FRD~\cite{DBLP:conf/eccv/ZyrianovZW22} (Frechet Range Distance) as a metric to measure the realism of synthetic data. The FRD score determines the quality of synthetic LiDAR point clouds by calculating the squared Wasserstein metric between the mean and covariance of network activations in RangeNet++~\cite{DBLP:conf/iros/MiliotoVBS19} for both synthetic and real data. The calculation of the FRD can be represented as follows:
\begin{equation}\footnotesize
F R D(X, Y)=\left\|\mu_X-\mu_Y\right\|^2+\operatorname{Tr}\left(\Sigma_X+\Sigma_Y-2\left(\Sigma_X \Sigma_Y\right)^{1 / 2}\right)
\vspace{-3pt}
\end{equation}
\noindent where $X$ and $Y$ represent the distributions of the network's activations for real and synthetic data, respectively. $\mu$ represents the means of the data activations, while $\Sigma$ is the covariance matrices of the activations from data. The trace of a matrix is denoted by $\operatorname{Tr}$. A lower FRD score indicates greater realism in the synthetic data.

\noindent\textbf{Baseline.} 
The V2X cooperative perception system is a complex software system involving multi-agent collaboration. However, the most advanced methods for manipulating objects in point clouds primarily focus on single-agent systems. Currently, the only available testing method for cooperative perception, CooTest~\cite{DBLP:conf/issta/GuoG00LGSF24}, does not support object manipulation within the scene. In order to evaluate the realism of the transformed data generated by V2X scene transformation operators in \tool~(i.e., RQ1), we collect corresponding point cloud data operators from state-of-the-art single-agent test generation methods as the baseline. Specifically, we leverage MultiTest~\cite{DBLP:conf/icse/GaoW000X24}, an object-level point cloud insertion-based test generation method, as the insertion operator. The deletion operator uses the method of directly deleting the object points in the ground-truth box. The translation, rotation, and scaling follow the methods proposed in LiRTest~\cite{DBLP:conf/issta/GuoF022}. 
Finally, we compare the realism of data generated by \tool~with that produced by combining these single-agent operators (denoted as \textit{SAGen}).
To evaluate the effectiveness of detecting errors in the cooperative perception system and retraining it for optimization~(i.e., RQ2 and RQ3), we use the only current cooperative perception testing method called CooTest~\cite{DBLP:conf/issta/GuoG00LGSF24} and a version of \tool~without a fitness-guided strategy (denoted as \tool\_N), which retains transformation data randomly.

\begin{table}[t]\footnotesize
  \centering
  \caption{Realism of transformed data generated across different approaches in diverse cooperative perception scenarios.}
  \vspace{-10pt}
  \renewcommand{\arraystretch}{0.75}
    \begin{tabular}{c|c|c|c|c|c|c|c|c|c|c|c|c|c}
    \hline
    \multicolumn{2}{c|}{\multirow{3}[6]{*}{\textbf{Scenario}}} & \multicolumn{6}{c|}{\textbf{V2V4Real}}        & \multicolumn{6}{c}{\textbf{V2XReal}} \bigstrut\\
\cline{3-14}
    \multicolumn{2}{c|}{} & \multicolumn{3}{c|}{\textbf{V2XGen}} & \multicolumn{3}{c|}{\textbf{SAGen}} & \multicolumn{3}{c|}{\textbf{V2XGen}} & \multicolumn{3}{c}{\textbf{SAGen}} \bigstrut\\
\cline{3-14}
    \multicolumn{2}{c|}{} & \textbf{M1} & \textbf{M2} & \textbf{M3} & \textbf{M1} & \textbf{M2} & \textbf{M3} & \textbf{M1} & \textbf{M2} & \textbf{M3} & \textbf{M1} & \textbf{M2} & \textbf{M3} \bigstrut\\
    \hline
    \multicolumn{2}{c|}{S1} & \textbf{202.8} & \textbf{205.6} & \textbf{211.3} & 223.4 & 247.8 & 280.9 & \textbf{14.3} & \textbf{25.0} & \textbf{33.9} & 14.9 & 26.8 & 34.6 \bigstrut\\
    \hline
    \multicolumn{2}{c|}{S2} & \textbf{192.0} & \textbf{200.5} & \textbf{209.0} & 201.7 & 236.7 & 286.8 & \textbf{18.3} & \textbf{43.6} & \textbf{63.0} & 49.9 & 69.6 & 80.2 \bigstrut\\
    \hline
    \multicolumn{2}{c|}{S3} & \textbf{241.5} & \textbf{260.3} & \textbf{294.9} & 252.7 & 304.2 & 328.3 & \textbf{18.6} & \textbf{27.8} & \textbf{51.5} & 39.7 & 50.0 & 62.6 \bigstrut\\
    \hline
    \multicolumn{2}{c|}{S4} & \textbf{171.7} & \textbf{184.0} & \textbf{196.9} & 177.7 & 218.7 & 232.1 & \textbf{24.4} & \textbf{44.4} & \textbf{65.8} & 43.4 & 49.8 & 69.4 \bigstrut\\
    \hline
    \multicolumn{2}{c|}{S5} & \textbf{170.0} & \textbf{198.5} & \textbf{255.7} & 188.1 & 232.9 & 281.0 & \textbf{34.1} & \textbf{64.5} & \textbf{88.6} & 106.8 & 123.2 & 141.0 \bigstrut\\
    \hline
    \multicolumn{2}{c|}{S6} & \textbf{140.2} & \textbf{147.5} & \textbf{156.0} & 151.2 & 166.4 & 187.4 & \textbf{38.3} & \textbf{66.7} & \textbf{99.4} & 67.1 & 81.3 & 101.4 \bigstrut\\
    \hline
    \multicolumn{2}{c|}{S7} & \textbf{121.8} & \textbf{149.5} & \textbf{168.3} & 132.9 & 171.6 & 199.1 & \textbf{16.5} & \textbf{28.6} & \textbf{46.1} & 22.3 & 35.3 & 50.3 \bigstrut\\
    \hline
    \multicolumn{2}{c|}{S8} & \textbf{203.1} & \textbf{215.4} & \textbf{219.7} & 220.0 & 254.0 & 284.9 & \textbf{15.5} & \textbf{36.4} & \textbf{43.3} & 50.0 & 59.7 & 67.1 \bigstrut\\
    \hline
    \multicolumn{2}{c|}{S9} & \textbf{236.3} & \textbf{282.7} & \textbf{378.3} & 247.3 & 320.2 & 450.4 & \textbf{42.4} & \textbf{76.0} & \textbf{98.8} & 78.1 & 93.6 & 111.8 \bigstrut\\
    \hline
    \multicolumn{2}{c|}{Average} & \textbf{186.6} & \textbf{204.9} & \textbf{232.2} & 199.5 & 239.2 & 281.2 & \textbf{24.7} & \textbf{45.9} & \textbf{65.6} & 52.5 & 65.5 & 79.8 \bigstrut\\
    \hline
    \end{tabular}%
    \label{tab:realism}
    \vspace{-5pt}  
\end{table}%

\begin{figure}[t]
	\centering
    \setlength{\abovecaptionskip}{0.1cm}
    \includegraphics[width=\linewidth, height=0.26\linewidth]{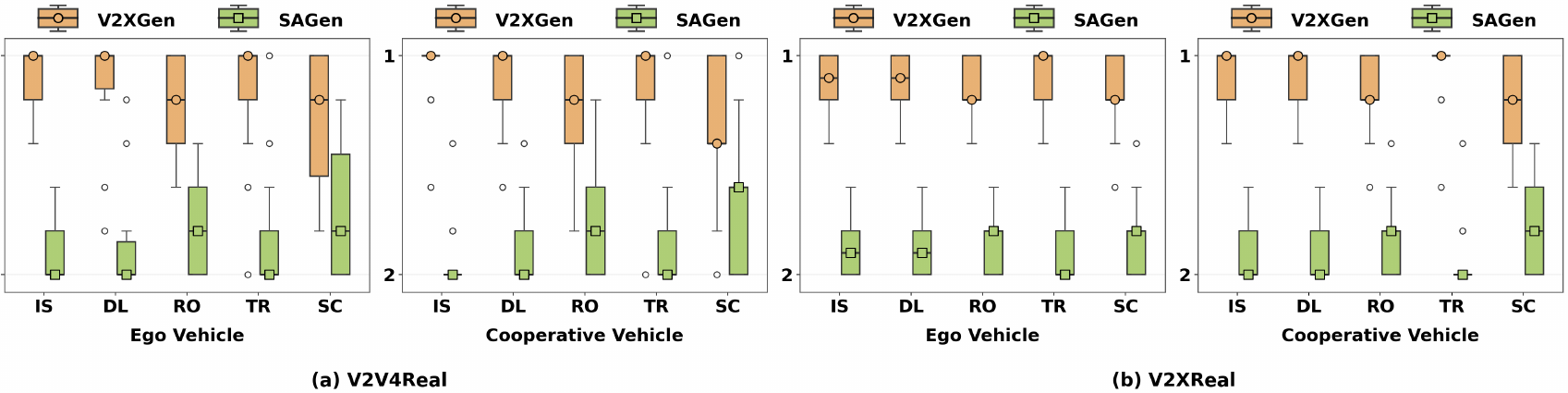}%
	\caption{Participants’ choices over different data synthesis methods in cooperative data naturalness.}
	\label{fig.rank}
    \vspace{-10pt}
\end{figure}

\section{Result Analysis}

\subsection{Answer to RQ1}

To assess the fidelity of the data generated by the operator in the \tool~framework, we conduct experiments utilizing all nine cooperative driving test scenarios (i.e., driving record for a period of time) from the V2V4Real test set. 
Specifically, we randomly select 50 scenes as the initial seed data for each scenario and perform 1, 2, and 3 transformations on each scene in the seed data, respectively, to form three synthetic data sets (denoted as $M_1$, $M_2$, and $M_3$). For each transformation, we randomly select an operator from the set $\mathbb{A}=\{IS, DL, TR, SC, RO\}$.
Throughout the data generation process, we ensure the consistency of the operators and coordinate transformations applied by both the ego vehicle and the cooperative vehicle. We then compute the FRD values of the \tool~and the baseline data generation method. The experiment is repeated five times, with results averaged to minimize randomness.

We further conduct a user study to evaluate the naturalness of cooperative driving data generated by each operator in \tool. We recruited 22 participants from the computer science departments of four research universities, with 95.5\% holding at least a master’s degree in SE or CS. Among them, 14 had over one year of experience in autonomous driving or V2X, and 17 had similar experience in object detection. During the study, we randomly selected five cooperative perception data instances, each containing both ego and cooperative vehicle data frames, as test seeds for each transformation operator. We then asked the participants to rank the cooperative perception data synthesized by two different generation methods from each seed through a questionnaire. Participants ranked the quality of synthesized point cloud data from two different generation methods for each seed, focusing on the realism of the ego and cooperative vehicle data frames. For each combination of transformation operator and viewpoint (ego or cooperative), each participant provided five rank values for each generation method, corresponding to the five test seeds. We calculated the arithmetic mean of these five rank values to represent each participant's overall ranking score for the corresponding generation method, transformation operator, and viewpoint. After the study, we used the one-sided Wilcoxon rank-sum test~\cite{cuzick1985wilcoxon} to assess whether any pipeline is significantly preferred by participants. To mitigate bias, we randomly assigned the order of data synthesized by different generation methods for each test seed. We further refer readers to our supplementary website~\cite{website} for the complete questionnaire.

\textbf{Results.} Table~\ref{tab:realism} shows the realism of quantitative results in diverse cooperative perception scenarios of different data synthesis pipelines. Compared with the baseline, \tool~achieves the lowest FRD in all configurations. On average, the \tool~method reduces FRD by 29.7\%, 22.1\%, and 17.6\% when compared to the SAGen method in the $M1$, $M2$, and $M3$ transformation configurations, respectively. These results indicate that \tool~can generate high-quality and relatively realistic cooperative perception data after the object transformation. Figure \ref{fig.rank} depicts the participants' preferences regarding two data synthesis strategies for each transformation operator in terms of the naturalness of the point cloud. The left and right sub-figures represent the synthesized data frames of the ego vehicle and cooperative vehicle, respectively, reflecting the realism achieved after applying the transformation operator. We find that more participants prefer \tool~over the baseline method on all five transformation operators. Wilcoxon’s rank-sum test results suggest that the mean ranking differences between \tool~and the second-chosen method are statistically significant (p\textless0.0001).

\textbf{Discussion.}
The results indicate that our V2X researchers in the field of autonomous driving generally agree that \tool~yields more realistic synthetic data than the single-agent method. These synthetic data adhere to the principles of the physical world and align with the characteristics of V2X systems. This conclusion supports the notion that the high-fidelity data generation method serves as a reasonable and effective approach for synthesizing complex V2X cooperative driving environments. Furthermore, FRD measures relative realism by quantifying the distance between the feature distributions of transformed synthetic data and real data. We observe that the FRD of the \tool~ is consistently lower than that of SAGen across all configurations. This result suggests that our generation method produces more realistic scenes than the baseline method. For both V2XGen and SAGen, each transformation introduces additional synthetic content into the scene. Consequently, the distributional distance between the transformed data and the real reference data progressively increases with the number of transformations. Additionally, we collect the experimental FRD results on the two datasets (Table~\ref{tab:realism}), together with the detection performance (AP) of the corresponding systems, and conduct a Spearman rank correlation analysis. Specifically, we calculate an overall Spearman rank correlation coefficient for each of the six cooperative perception models using the paired FRD and AP. These results indicate a significant negative association ($\rho$<0, $p$<0.05) between FRD and AP across all six models. The results show that no statistically significant positive correlation between FRD and AP can be inferred. Therefore, the empirical results provide no evidence that the realistic scenarios generated by V2XGen introduce unrealistic perturbations that systematically reduce detection performance.

\subsection{Answer to RQ2}
To evaluate the effectiveness of the proposed fitness-guided V2X scene generation strategy in detecting erroneous behaviors of V2X cooperative perception systems, we randomly split the original test set into two equal subsets, denoted as $T_{h1}$ and $T_{h2}$, which correspond to a 50\% and 50\% partition. We leverage one half (denoted as $T_{h1}$) as the initial seeds. To reduce randomness and cultivate diverse synthetic data, we set $N$ to 1, 2, and 3 in Algorithm~\ref{alg1} to generate three distinct transformed datasets. We then combined these datasets to create a larger transformed test set.
In the experiments, we apply \tool, CooTest, and \tool\_N (w/o fitness-guided strategy) to select a specific small batch size from the generated data, enabling a comparison of the effectiveness of different guided methods in test generation. Specifically, for each guided strategy, we retain 10\% and 15\% of the transformed data to evaluate the effectiveness of the guidance strategies.
Then we calculate the AP in generated test cases of cooperative perception systems under test for each configuration. Furthermore, we count the number of long-range and occlusion perception errors (see Section~\ref{Definitions}), the occlusion perception error rate~(OER), and the long-distance perception error rate~(LER), and visualize some typical instances of long-range perception errors and occlusion perception errors detected by \tool~to provide a direct understanding. To examine the impact of different configurations of the hyperparameters $\alpha$ and $\beta$ on the number of occlusion and long-range perception errors, we conduct comparative experiments using two extreme configurations, where $\alpha = 1, \beta = 0$ corresponds to the occlusion-only configuration, while $\alpha = 0, \beta = 1$ corresponds to the long-range-only configuration. The experiment is repeated three times and averaged to minimize the effect of randomness.

\begin{table}[t!]
  \centering
  \caption{Testing results of cooperative perception systems on the V2V4Real transformed testing sets generated with different guidance approaches.}
  \vspace{-5pt}
  \label{tab:rq2sel-v2v4real}
  \footnotesize
  \setlength{\tabcolsep}{5pt}
  \begin{tabular}{l *{3}{c} *{3}{c} *{3}{c}}
    \toprule
    \multirow{2}{*}{\textbf{System}}
      & \multicolumn{3}{c}{\textbf{V2XGen}}
      & \multicolumn{3}{c}{\textbf{CooTest}}
      & \multicolumn{3}{c}{\textbf{Random}} \\
    \cmidrule(lr){2-4}\cmidrule(lr){5-7}\cmidrule(lr){8-10}
      & \textbf{AP} & \textbf{OE} & \textbf{LE}
      & \textbf{AP} & \textbf{OE} & \textbf{LE}
      & \textbf{AP} & \textbf{OE} & \textbf{LE} \\
    \midrule
    Early Fusion & \textbf{54.1\pmm{0.4}} & \textbf{1887\pmm{39}} & \textbf{672\pmm{31}}  & 60.6\pmm{1.3} & 626\pmm{103} & 361\pmm{22} & 59.6\pmm{0.1} & 653\pmm{39} & 399\pmm{23} \\
    Late Fusion  & \textbf{45.0\pmm{0.1}} & \textbf{1562\pmm{25}} & \textbf{604\pmm{26}}  & 53.1\pmm{2.3} & 386\pmm{32}  & 236\pmm{8}  & 55.1\pmm{0.5} & 379\pmm{8}  & 246\pmm{16} \\
    V2VNet       & \textbf{51.6\pmm{0.2}} & \textbf{1656\pmm{46}} & \textbf{1018\pmm{25}} & 64.1\pmm{0.8} & 488\pmm{57}  & 352\pmm{34} & 64.2\pmm{0.0} & 482\pmm{32} & 367\pmm{17} \\
    V2X-ViT      & \textbf{52.4\pmm{0.3}} & \textbf{1927\pmm{39}} & \textbf{962\pmm{30}}  & 64.5\pmm{1.5} & 530\pmm{53}  & 396\pmm{33} & 64.5\pmm{0.6} & 520\pmm{18} & 438\pmm{4}  \\
    AttFusion    & \textbf{52.8\pmm{0.3}} & \textbf{1695\pmm{31}} & \textbf{855\pmm{17}}  & 63.7\pmm{1.0} & 465\pmm{52}  & 383\pmm{23} & 64.5\pmm{1.0} & 429\pmm{48} & 396\pmm{20} \\
    F-Cooper     & \textbf{48.3\pmm{0.2}} & \textbf{2012\pmm{32}} & \textbf{1452\pmm{25}} & 60.0\pmm{1.2} & 620\pmm{57}  & 579\pmm{52} & 60.4\pmm{0.4} & 637\pmm{24} & 607\pmm{58} \\
    \bottomrule
  \end{tabular}
  \vspace{-5pt}
\end{table}

\begin{table}[t!]
  \centering
  \caption{Testing results of cooperative perception systems on the V2XReal transformed testing sets generated with different guidance approaches.}
  \vspace{-5pt}
  \label{tab:rq2sel-V2Xreal}
  \footnotesize
  \setlength{\tabcolsep}{5pt}
  \begin{tabular}{l *{3}{c} *{3}{c} *{3}{c}}
    \toprule
    \multirow{2}{*}{\textbf{System}}
      & \multicolumn{3}{c}{\textbf{V2XGen}}
      & \multicolumn{3}{c}{\textbf{CooTest}}
      & \multicolumn{3}{c}{\textbf{Random}} \\
    \cmidrule(lr){2-4}\cmidrule(lr){5-7}\cmidrule(lr){8-10}
      & \textbf{AP} & \textbf{OE} & \textbf{LE}
      & \textbf{AP} & \textbf{OE} & \textbf{LE}
      & \textbf{AP} & \textbf{OE} & \textbf{LE} \\
    \midrule
    Early Fusion
      & \textbf{44.0\pmm{0.3}} & \textbf{668\pmm{5}} & \textbf{1236\pmm{8}}
      & 58.7\pmm{0.5} & 326\pmm{8} & 841\pmm{9}
      & 52.4\pmm{0.3} & 345\pmm{3} & 945\pmm{19} \\

    Late Fusion
      & \textbf{43.2\pmm{0.1}} & \textbf{675\pmm{10}} & \textbf{1218\pmm{2}}
      & 55.7\pmm{0.3} & 358\pmm{14} & 811\pmm{6}
      & 50.5\pmm{0.1} & 344\pmm{4} & 959\pmm{15} \\

    V2VNet
      & \textbf{47.1\pmm{0.1}} & \textbf{565\pmm{9}} & \textbf{1194\pmm{4}}
      & 58.6\pmm{0.3} & 262\pmm{4} & 865\pmm{19}
      & 55.8\pmm{0.4} & 268\pmm{6} & 894\pmm{18} \\

    V2X-ViT
      & \textbf{29.6\pmm{0.1}} & \textbf{981\pmm{4}} & \textbf{1275\pmm{9}}
      & 43.7\pmm{0.3} & 578\pmm{4} & 957\pmm{17}
      & 36.5\pmm{0.9} & 561\pmm{9} & 1067\pmm{42} \\

    AttFusion
      & \textbf{36.2\pmm{0.3}} & \textbf{872\pmm{11}} & \textbf{1308\pmm{1}}
      & 52.0\pmm{0.1} & 502\pmm{8} & 900\pmm{7}
      & 43.0\pmm{0.3} & 486\pmm{2} & 1042\pmm{24} \\

    F-Cooper
      & \textbf{41.9\pmm{0.1}} & \textbf{666\pmm{10}} & \textbf{1278\pmm{15}}
      & 54.6\pmm{0.3} & 343\pmm{4} & 829\pmm{8}
      & 49.8\pmm{0.4} & 335\pmm{9} & 970\pmm{24} \\
    \bottomrule
  \end{tabular}
  \vspace{-10pt}
\end{table}

\textbf{Results.}
Table~\ref{tab:rq2sel-v2v4real} and Table~\ref{tab:rq2sel-V2Xreal} report the testing performance of different systems on test scenes generated using three guidance approaches for the V2V4Real and V2XReal datasets, respectively. Compared to the \tool\_N and the CooTest guidance strategy, the proposed fitness-guided V2X scene generation method can find more occlusion and long-range perception errors in all configurations.
When retaining 15\% of the transformation data, the fitness-guided V2X scenario generation strategy exhibits an average effectiveness improvement of 9.2\%, respectively, compared to the best baseline performance. Furthermore, experimental results show that a large number of occlusion and long-range perception errors could lead to a sharp decline in AP. These results underscore that \tool~can achieve higher testing efficiency by using the fitness-guided V2X scene generation method. Table~\ref{tab:oer_ler-v2v4real} and Table~\ref{tab:oer_ler-v2xreal} present the occlusion and long-range perception error rates of the system evaluated on data generated using various guidance methods for the V2V4Real and V2XReal datasets, respectively. Results show that V2XGen-guided data produces data with higher occlusion error rates and higher long-range perception error rates compared to other baselines. Moreover, we provide detailed results for the 10\% retaining data configuration, and several visualizations of occlusion and long-range errors on the supplementary website~\cite{website}.
Table~\ref{tab:hyperparameters} illustrates the effects of different hyperparameter configurations on the number of occlusion and long-range perception errors. As shown in the table, assigning a higher weight to $F_{OP}$ in Equation~\ref{equa5} increases the number of detected occlusion perception errors, whereas assigning a higher weight to $F_{LP}$ increases the number of detected long-range perception errors.

\begin{table}[t!]
  \centering
  \caption{The occlusion and long-range perception error rates of V2V4Real cooperative perception test data generated through various guidance methods.}
  \vspace{-5pt}
  \label{tab:oer_ler-v2v4real}
  \footnotesize
  \setlength{\tabcolsep}{6pt}
  \begin{tabular}{l *{2}{c} *{2}{c} *{2}{c}}
    \toprule
    \multirow{2}{*}{\textbf{System}}
      & \multicolumn{2}{c}{\textbf{V2XGen}}
      & \multicolumn{2}{c}{\textbf{CooTest}}
      & \multicolumn{2}{c}{\textbf{Random}} \\
    \cmidrule(lr){2-3}\cmidrule(lr){4-5}\cmidrule(lr){6-7}
      & \textbf{OER} & \textbf{LER}
      & \textbf{OER} & \textbf{LER}
      & \textbf{OER} & \textbf{LER} \\
    \midrule
    Early Fusion    & \textbf{0.37\pmm{0.01}} & \textbf{0.28\pmm{0.01}} & 0.27\pmm{0.02} & 0.18\pmm{0.01} & 0.27\pmm{0.01} & 0.19\pmm{0.01} \\
    Late Fusion     & \textbf{0.31\pmm{0.01}} & \textbf{0.28\pmm{0.00}} & 0.17\pmm{0.01} & 0.14\pmm{0.01} & 0.16\pmm{0.01} & 0.13\pmm{0.01} \\
    V2VNet    & \textbf{0.35\pmm{0.01}} & \textbf{0.29\pmm{0.00}} & 0.21\pmm{0.00} & 0.18\pmm{0.01} & 0.21\pmm{0.02} & 0.17\pmm{0.01} \\
    V2X-ViT   & \textbf{0.37\pmm{0.00}} & \textbf{0.32\pmm{0.00}} & 0.23\pmm{0.01} & 0.20\pmm{0.01} & 0.22\pmm{0.01} & 0.20\pmm{0.01} \\
    AttFusion & \textbf{0.35\pmm{0.01}} & \textbf{0.26\pmm{0.01}} & 0.20\pmm{0.01} & 0.19\pmm{0.01} & 0.18\pmm{0.02} & 0.19\pmm{0.01} \\
    F-Cooper  & \textbf{0.45\pmm{0.01}} & \textbf{0.40\pmm{0.01}} & 0.27\pmm{0.01} & 0.29\pmm{0.02} & 0.27\pmm{0.00} & 0.30\pmm{0.01} \\
    \bottomrule
  \end{tabular}
  \vspace{-5pt}
\end{table}

\begin{table}[t!]
  \centering
  \caption{The occlusion and long-range perception error rates of V2XReal cooperative perception test data generated through various guidance methods.}
  \vspace{-5pt}
  \label{tab:oer_ler-v2xreal}
  \footnotesize
  \setlength{\tabcolsep}{6pt}
  \begin{tabular}{l *{2}{c} *{2}{c} *{2}{c}}
    \toprule
    \multirow{2}{*}{\textbf{System}}
      & \multicolumn{2}{c}{\textbf{V2XGen}}
      & \multicolumn{2}{c}{\textbf{CooTest}}
      & \multicolumn{2}{c}{\textbf{V2XGen\_N}} \\
    \cmidrule(lr){2-3}\cmidrule(lr){4-5}\cmidrule(lr){6-7}
      & \textbf{OER} & \textbf{LER}
      & \textbf{OER} & \textbf{LER}
      & \textbf{OER} & \textbf{LER} \\
    \midrule
    Early Fusion
      & \textbf{0.58\pmm{0.00}} & \textbf{0.84\pmm{0.00}}
      & 0.32\pmm{0.01} & 0.68\pmm{0.00}
      & 0.40\pmm{0.00} & 0.78\pmm{0.01} \\

    Late Fusion
      & \textbf{0.58\pmm{0.00}} & \textbf{0.85\pmm{0.00}}
      & 0.34\pmm{0.01} & 0.68\pmm{0.01}
      & 0.40\pmm{0.01} & 0.81\pmm{0.02} \\

    V2VNet
      & \textbf{0.52\pmm{0.00}} & \textbf{0.82\pmm{0.00}}
      & 0.27\pmm{0.00} & 0.65\pmm{0.01}
      & 0.32\pmm{0.01} & 0.73\pmm{0.01} \\

    V2X-ViT
      & \textbf{0.78\pmm{0.01}} & \textbf{0.90\pmm{0.00}}
      & 0.57\pmm{0.00} & 0.79\pmm{0.01}
      & 0.65\pmm{0.01} & 0.89\pmm{0.00} \\

    AttFusion
      & \textbf{0.71\pmm{0.01}} & \textbf{0.90\pmm{0.00}}
      & 0.46\pmm{0.00} & 0.76\pmm{0.00}
      & 0.57\pmm{0.01} & 0.87\pmm{0.01} \\

    F-Cooper
      & \textbf{0.58\pmm{0.01}} & \textbf{0.85\pmm{0.00}}
      & 0.33\pmm{0.01} & 0.72\pmm{0.00}
      & 0.39\pmm{0.01} & 0.79\pmm{0.02} \\
    \bottomrule
  \end{tabular}
  \vspace{-5pt}
\end{table}

\begin{table}[t!]\scriptsize
  \centering
  \caption{The impact of different configurations of hyperparameters $\alpha$ and $\beta$ on the number of occlusion and long-range perception errors.}
    \vspace{-5pt}
    \begin{tabular}{c|c|c|c|c|c|c|c|c}
    \hline
    \multirow{3}[6]{*}{\textbf{System}} 
    & \multicolumn{4}{c|}{\textbf{V2V4Real}} 
    & \multicolumn{4}{c}{\textbf{V2XReal}} \bigstrut\\
\cline{2-9}          
    & \multicolumn{2}{c|}{\textbf{$\alpha$=1, $\beta$=0}} 
    & \multicolumn{2}{c|}{\textbf{$\alpha$=0, $\beta$=1}} 
    & \multicolumn{2}{c|}{\textbf{$\alpha$=1, $\beta$=0}} 
    & \multicolumn{2}{c}{\textbf{$\alpha$=0, $\beta$=1}} \bigstrut\\
\cline{2-9}          
    & \textbf{OE} & \textbf{LE} 
    & \textbf{OE} & \textbf{LE} 
    & \textbf{OE} & \textbf{LE} 
    & \textbf{OE} & \textbf{LE} \bigstrut\\
    \hline
    Early Fusion 
    & 1833 ± 17 & 464 ± 10 
    & 1420 ± 11 & 1056 ± 11 
    & 729 ± 11 & 834 ± 9 
    & 459 ± 32 & 1345 ± 1 \bigstrut\\
    \hline
    Late Fusion 
    & 1523 ± 24 & 370 ± 23 
    & 1308 ± 13 & 805 ± 7 
    & 732 ± 6 & 854 ± 3 
    & 474 ± 22 & 1365 ± 7 \bigstrut\\
    \hline
    V2VNet 
    & 1706 ± 31 & 417 ± 35 
    & 1185 ± 32 & 1333 ± 5 
    & 631 ± 8 & 774 ± 15 
    & 382 ± 29 & 1316 ± 11 \bigstrut\\
    \hline
    V2X-ViT 
    & 2073 ± 56 & 497 ± 34 
    & 1312 ± 23 & 1421 ± 12 
    & 1052 ± 32 & 894 ± 17 
    & 692 ± 24 & 1498 ± 2 \bigstrut\\
    \hline
    AttFusion 
    & 1690 ± 20 & 497 ± 19 
    & 1302 ± 20 & 1144 ± 12 
    & 956 ± 18 & 923 ± 23 
    & 651 ± 17 & 1439 ± 17 \bigstrut\\
    \hline
    F-Cooper 
    & 2296 ± 17 & 715 ± 47 
    & 1509 ± 32 & 1815 ± 5 
    & 702 ± 6 & 874 ± 4 
    & 458 ± 25 & 1405 ± 3 \bigstrut\\
    \hline
    \end{tabular}%
    \vspace{-10pt}
  \label{tab:hyperparameters}%
\end{table}

\textbf{Discussion.}
We analyze the relationship between AP performance and the occlusion and long-range perception errors identified by scene generation methods under different guidance strategies. We compute the Spearman rank correlation coefficients between AP and the error metrics (OE and LE) across all system and guidance configurations. The results demonstrate that AP is negatively correlated with both OE and LE ($\rho < 0$), and these correlations are statistically significant ($p < 0.05$).
Moreover, to evaluate the contribution of each of the five LiDAR mutation operators to detection performance, we assess the individual effects of each operator. We use $Th_1$ as the initial seed set. For each initial seed in $Th_1$, we apply each mutation operator once. We then calculate the average detection performance (AP) across all models for each operator on different datasets. Across all datasets, the results show that the insertion operator has the smallest effect on detection performance, whereas the rotation operator has the largest effect. The greater impact of the rotation operator can be attributed to its alteration of object orientations, which poses greater challenges to cooperative perception systems in detecting objects at certain orientations. Furthermore, our definitions of occlusion and long-range perception errors are rooted in the fundamental principle of V2X cooperative perception systems, which posits that such systems should assist the ego vehicle in detecting occlusions and distant objects. However, existing systems can only partially mitigate the challenges posed by individual agents.

\subsection{Answer to RQ3}

Here we investigate whether retraining the cooperative perception systems with test scenes generated by \tool~helps enhance the system's performance. 
We apply the generated test suite (derived from $T_{h1}$) in RQ2 for system retraining. We keep the experimental configuration of retraining consistent with the initial training system process, and the retraining epoch is set as 10. 
To demonstrate the improvement of performance clearly, we utilize $T_{h2}$ as the initial seed and set $N$ to 1, 2, and 3 to execute Algorithm~\ref{alg1} for generating three distinct transformed datasets. Subsequently, we combine all these transformed datasets to create a large-scale validation set.
To further ensure that the retrained systems perform effectively in the original scenes, we test all retrained systems on the original test data $T_{h2}$. The experiment is repeated three times and averaged to minimize the effect of randomness.

\begin{table}[t!]
  \centering
  \caption{The test result comparison on the V2V4Real transformed dataset after retraining the system with the transformed data.}
   \vspace{-5pt}
  \label{tab:transformed-v2v4real}
  \scriptsize
  \setlength{\tabcolsep}{2.5pt}
  \begin{tabular}{l *{4}{c} *{4}{c} *{4}{c}}
    \toprule
    \multirow{3}{*}{\textbf{System}}
      & \multicolumn{4}{c}{\textbf{AP}}
      & \multicolumn{4}{c}{\textbf{OE}}
      & \multicolumn{4}{c}{\textbf{LE}} \\
    \cmidrule(lr){2-5}\cmidrule(lr){6-9}\cmidrule(lr){10-13}
      & \multirow{2}{*}{\textbf{Before}} & \multicolumn{3}{c}{\textbf{After}}
      & \multirow{2}{*}{\textbf{Before}} & \multicolumn{3}{c}{\textbf{After}}
      & \multirow{2}{*}{\textbf{Before}} & \multicolumn{3}{c}{\textbf{After}} \\
    \cmidrule(lr){3-5}\cmidrule(lr){7-9}\cmidrule(lr){11-13}
      &       & \textbf{V2XGen} & \textbf{CooTest} & \textbf{Random}
      &       & \textbf{V2XGen} & \textbf{CooTest} & \textbf{Random}
      &       & \textbf{V2XGen} & \textbf{CooTest} & \textbf{Random} \\
    \midrule
    Early Fusion    & 58.2 & \textbf{64.7\pmm{0.9}} & 58.8\pmm{0.5} & 61.3\pmm{1.4} & 2696 & \textbf{1494\pmm{77}}  & 2565\pmm{93}  & 2185\pmm{102} & 1491 & \textbf{929\pmm{26}}   & 1448\pmm{91}  & 1294\pmm{153} \\
    Late Fusion     & 59.3 & \textbf{69.4\pmm{0.3}} & 66.5\pmm{0.1} & 68.4\pmm{0.2} & 1631 & \textbf{1151\pmm{16}}  & 1743\pmm{54}  & 1396\pmm{44}  & 829  & \textbf{824\pmm{14}}   & 918\pmm{4}    & 865\pmm{17}   \\
    V2VNet    & 61.0 & \textbf{68.7\pmm{0.1}} & 62.4\pmm{1.2} & 64.7\pmm{2.1} & 2259 & \textbf{1175\pmm{31}}  & 2182\pmm{49}  & 1655\pmm{114} & 1581 & \textbf{1137\pmm{24}}  & 1614\pmm{89}  & 1477\pmm{159} \\
    V2X-ViT   & 61.4 & \textbf{71.3\pmm{3.5}} & 67.6\pmm{3.1} & 67.7\pmm{2.8} & 2254 & \textbf{894\pmm{228}}  & 1155\pmm{167} & 1128\pmm{186} & 1730 & \textbf{990\pmm{395}}  & 1224\pmm{165} & 1198\pmm{241} \\
    AttFusion & 62.6 & \textbf{74.0\pmm{0.7}} & 66.5\pmm{3.3} & 68.7\pmm{1.5} & 2068 & \textbf{865\pmm{16}}   & 1654\pmm{357} & 1239\pmm{230} & 1495 & \textbf{1088\pmm{117}} & 1302\pmm{121} & 1189\pmm{335} \\
    F-Cooper  & 58.2 & \textbf{67.1\pmm{0.1}} & 64.7\pmm{0.8} & 64.4\pmm{0.8} & 2745 & \textbf{1117\pmm{7}}   & 1622\pmm{41}  & 1552\pmm{10}  & 2325 & \textbf{1030\pmm{17}}  & 1262\pmm{9}   & 1241\pmm{45}  \\
    \bottomrule
  \end{tabular}
  \vspace{-10pt}
\end{table}

\begin{table}[t!]

  \centering

  \caption{The test result comparison on the V2XReal transformed dataset after retraining the system with the transformed data.}
  \label{tab:transformed-v2Xreal}
  \vspace{-5pt}
  \scriptsize

  \setlength{\tabcolsep}{2.5pt}

  \begin{tabular}{l *{4}{c} *{4}{c} *{4}{c}}

    \toprule

    \multirow{3}{*}{\textbf{System}}

      & \multicolumn{4}{c}{\textbf{AP}}

      & \multicolumn{4}{c}{\textbf{OE}}

      & \multicolumn{4}{c}{\textbf{LE}} \\

    \cmidrule(lr){2-5}\cmidrule(lr){6-9}\cmidrule(lr){10-13}

      & \multirow{2}{*}{\textbf{Before}} & \multicolumn{3}{c}{\textbf{After}}

      & \multirow{2}{*}{\textbf{Before}} & \multicolumn{3}{c}{\textbf{After}}

      & \multirow{2}{*}{\textbf{Before}} & \multicolumn{3}{c}{\textbf{After}} \\

    \cmidrule(lr){3-5}\cmidrule(lr){7-9}\cmidrule(lr){11-13}

      &       & \textbf{V2XGen} & \textbf{CooTest} & \textbf{Random}

      &       & \textbf{V2XGen} & \textbf{CooTest} & \textbf{Random}

      &       & \textbf{V2XGen} & \textbf{CooTest} & \textbf{Random} \\

    \midrule

    Early Fusion

      & 65.9

      & \textbf{69.7\pmm{0.5}}

      & 67.1\pmm{0.7}

      & 67.8\pmm{0.9}

      & 464

      & \textbf{358\pmm{8}}

      & 427\pmm{14}

      & 389\pmm{18}

      & 1474

      & \textbf{1333\pmm{36}}

      & 1414\pmm{36}

      & 1394\pmm{48} \\

    Late Fusion

      & 61.6

      & \textbf{71.5\pmm{0.4}}

      & 70.4\pmm{0.9}

      & 70.2\pmm{0.6}

      & 375

      & \textbf{305\pmm{11}}

      & 329\pmm{33}

      & 333\pmm{28}

      & 1380

      & \textbf{1130\pmm{52}}

      & 1243\pmm{33}

      & 1245\pmm{29} \\

    V2VNet

      & 62.3

      & \textbf{70.3\pmm{2.1}}

      & 67.8\pmm{0.4}

      & 68.6\pmm{0.7}

      & 501

      & \textbf{376\pmm{24}}

      & 421\pmm{11}

      & 443\pmm{8}

      & 1314

      & \textbf{1114\pmm{68}}

      & 1278\pmm{43}

      & 1240\pmm{34} \\

    V2X-ViT

      & 42.8

      & \textbf{60.6\pmm{1.2}}

      & 54.9\pmm{1.2}

      & 57.7\pmm{0.5}

      & 924

      & \textbf{493\pmm{32}}

      & 689\pmm{87}

      & 672\pmm{48}

      & 1774

      & \textbf{1435\pmm{41}}

      & 1646\pmm{77}

      & 1637\pmm{63} \\

    AttFusion

      & 50.9

      & \textbf{67.1\pmm{0.9}}

      & 64.3\pmm{1.9}

      & 62.6\pmm{1.1}

      & 794

      & \textbf{499\pmm{77}}

      & 583\pmm{18}

      & 605\pmm{35}

      & 1720

      & \textbf{1326\pmm{98}}

      & 1474\pmm{56}

      & 1475\pmm{62} \\

    F-Cooper

      & 64.5

      & \textbf{70.9\pmm{0.4}}

      & 69.2\pmm{0.4}

      & 68.4\pmm{1.1}

      & 395

      & \textbf{330\pmm{25}}

      & 359\pmm{10}

      & 382\pmm{19}

      & 1561

      & \textbf{1174\pmm{21}}

      & 1329\pmm{60}

      & 1333\pmm{59} \\

    \bottomrule

  \end{tabular}
  \vspace{-15pt}
\end{table}

\textbf{Results.}
Table~\ref{tab:transformed-v2v4real} to Table \ref{tab:original-V2XReal} present the AP, occlusion perception error numbers, and long-range perception error numbers of all systems after retraining with the transformed data. It is evident that, regardless of the fusion scheme employed by the cooperative perception system, \tool~can consistently enhance the AP on the original test set and the transformed test set and mitigate occlusion perception error and long-range perception error after retraining. Furthermore, the performance achieved by \tool~through retraining across all configurations surpasses that of the baselines. In comparison to the perception performance under the transformed test set before retraining, the overall improvement in AP ranges from 3.8\% to 17.8\%. Additionally, the average reductions in occlusion and long-range perception errors are 38.9\% and 25.1\%, respectively. Furthermore, the average precision of all retrained systems evaluated on the original test scenes increases by 9.1\%.

\begin{table}[t!]
  \centering
  \caption{The test result comparison on the V2V4Real original dataset after retraining the system with the transformed data.}
  \label{tab:original-v2v4real}
  \vspace{-5pt}
  \scriptsize
  \setlength{\tabcolsep}{2.5pt}
  \begin{tabular}{l *{4}{c} *{4}{c} *{4}{c}}
    \toprule
    \multirow{3}{*}{\textbf{System}}
      & \multicolumn{4}{c}{\textbf{AP}}
      & \multicolumn{4}{c}{\textbf{OE}}
      & \multicolumn{4}{c}{\textbf{LE}} \\
    \cmidrule(lr){2-5}\cmidrule(lr){6-9}\cmidrule(lr){10-13}
      & \multirow{2}{*}{\textbf{Before}} & \multicolumn{3}{c}{\textbf{After}}
      & \multirow{2}{*}{\textbf{Before}} & \multicolumn{3}{c}{\textbf{After}}
      & \multirow{2}{*}{\textbf{Before}} & \multicolumn{3}{c}{\textbf{After}} \\
    \cmidrule(lr){3-5}\cmidrule(lr){7-9}\cmidrule(lr){11-13}
      &       & \textbf{V2XGen} & \textbf{CooTest} & \textbf{Random}
      &       & \textbf{V2XGen} & \textbf{CooTest} & \textbf{Random}
      &       & \textbf{V2XGen} & \textbf{CooTest} & \textbf{Random} \\
    \midrule
    Early Fusion    & 59.4 & \textbf{64.5\pmm{0.9}} & 60.0\pmm{0.8} & 62.5\pmm{1.5} & 1690 & \textbf{974\pmm{48}}  & 1603\pmm{78}  & 1348\pmm{50}  & 902  & \textbf{566\pmm{16}}  & 877\pmm{63}   & 796\pmm{112}  \\
    Late Fusion     & 55.2 & \textbf{68.4\pmm{0.4}} & 66.6\pmm{0.2} & 68.1\pmm{0.3} & 953  & \textbf{685\pmm{7}}   & 1018\pmm{33}  & 826\pmm{26}   & 560  & \textbf{563\pmm{6}}   & 612\pmm{6}    & 584\pmm{9}    \\
    V2VNet    & 64.4 & \textbf{70.5\pmm{0.2}} & 66.7\pmm{1.8} & 68.0\pmm{2.2} & 1251 & \textbf{619\pmm{14}}  & 1191\pmm{29}  & 911\pmm{73}   & 865  & \textbf{665\pmm{13}}  & 887\pmm{69}   & 814\pmm{106}  \\
    V2X-ViT   & 64.7 & \textbf{73.4\pmm{3.1}} & 71.1\pmm{3.9} & 70.4\pmm{3.2} & 1275 & \textbf{487\pmm{136}} & 635\pmm{132}  & 639\pmm{132}  & 971  & \textbf{576\pmm{241}} & 677\pmm{116}  & 686\pmm{158}  \\
    AttFusion & 65.0 & \textbf{74.6\pmm{0.8}} & 69.1\pmm{3.2} & 70.8\pmm{1.3} & 1137 & \textbf{458\pmm{2}}   & 916\pmm{197}  & 671\pmm{131}  & 912  & \textbf{737\pmm{87}}  & 785\pmm{82}   & 714\pmm{208}  \\
    F-Cooper  & 60.7 & \textbf{68.8\pmm{0.2}} & 66.7\pmm{0.8} & 66.3\pmm{0.8} & 1602 & \textbf{619\pmm{6}}   & 905\pmm{25}   & 870\pmm{4}    & 1416 & \textbf{623\pmm{10}}  & 773\pmm{10}   & 771\pmm{26}   \\
    \bottomrule
  \end{tabular}
  \vspace{-10pt}
\end{table}

\begin{table}[t!]
  \centering
  \caption{The test result comparison on the V2XReal original dataset after retraining the system with the transformed data.}
  \label{tab:original-V2XReal}
   \vspace{-5pt}
  \scriptsize
  \setlength{\tabcolsep}{2.5pt}
  \begin{tabular}{l *{4}{c} *{4}{c} *{4}{c}}
    \toprule
    \multirow{3}{*}{\textbf{System}}
      & \multicolumn{4}{c}{\textbf{AP}}
      & \multicolumn{4}{c}{\textbf{OE}}
      & \multicolumn{4}{c}{\textbf{LE}} \\
    \cmidrule(lr){2-5}\cmidrule(lr){6-9}\cmidrule(lr){10-13}
      & \multirow{2}{*}{\textbf{Before}} & \multicolumn{3}{c}{\textbf{After}}
      & \multirow{2}{*}{\textbf{Before}} & \multicolumn{3}{c}{\textbf{After}}
      & \multirow{2}{*}{\textbf{Before}} & \multicolumn{3}{c}{\textbf{After}} \\
    \cmidrule(lr){3-5}\cmidrule(lr){7-9}\cmidrule(lr){11-13}
      &       & \textbf{V2XGen} & \textbf{CooTest} & \textbf{Random}
      &       & \textbf{V2XGen} & \textbf{CooTest} & \textbf{Random}
      &       & \textbf{V2XGen} & \textbf{CooTest} & \textbf{Random} \\
    \midrule
    Early Fusion
      & 56.9
      & \textbf{62.7\pmm{0.7}}
      & 60.7\pmm{0.9}
      & 61.6\pmm{0.3}
      & 515
      & \textbf{437\pmm{16}}
      & 497\pmm{41}
      & 463\pmm{11}
      & 2314
      & \textbf{2112\pmm{70}}
      & 2248\pmm{51}
      & 2202\pmm{43} \\

    Late Fusion
      & 54.9
      & \textbf{65.3\pmm{1.1}}
      & 62.1\pmm{0.9}
      & 63.1\pmm{1.8}
      & 534
      & \textbf{389\pmm{23}}
      & 426\pmm{26}
      & 425\pmm{31}
      & 2349
      & \textbf{1926\pmm{191}}
      & 2118\pmm{23}
      & 2063\pmm{97} \\

    V2VNet
      & 59.9
      & \textbf{65.3\pmm{1.5}}
      & 62.8\pmm{0.6}
      & 63.2\pmm{1.0}
      & 457
      & \textbf{347\pmm{46}}
      & 413\pmm{24}
      & 397\pmm{36}
      & 2253
      & \textbf{1957\pmm{188}}
      & 2133\pmm{57}
      & 2066\pmm{158} \\

    V2X-ViT
      & 40.3
      & \textbf{58.8\pmm{1.8}}
      & 52.5\pmm{3.6}
      & 55.3\pmm{1.3}
      & 938
      & \textbf{544\pmm{107}}
      & 686\pmm{105}
      & 681\pmm{77}
      & 2850
      & \textbf{2164\pmm{308}}
      & 2616\pmm{188}
      & 2428\pmm{266} \\

    AttFusion
      & 47.0
      & \textbf{61.5\pmm{0.9}}
      & 59.3\pmm{1.0}
      & 58.5\pmm{1.4}
      & 806
      & \textbf{505\pmm{10}}
      & 595\pmm{91}
      & 562\pmm{71}
      & 2759
      & \textbf{2019\pmm{287}}
      & 2349\pmm{60}
      & 2305\pmm{271} \\

    F-Cooper
      & 55.0
      & \textbf{64.9\pmm{1.6}}
      & 62.4\pmm{0.6}
      & 62.7\pmm{0.9}
      & 469
      & \textbf{361\pmm{29}}
      & 411\pmm{20}
      & 416\pmm{20}
      & 2525
      & \textbf{1931\pmm{218}}
      & 2229\pmm{150}
      & 2106\pmm{204} \\

    \bottomrule
  \end{tabular}
   \vspace{-15pt}
\end{table}

\textbf{Discussion.}
Our experiments show that \tool~reduces occlusion perception errors and long-range perception errors by leveraging retraining techniques to increase the cooperative perception system's overall performance. By using \tool~to generate realistic transformation scenes based on the collected real scenes, the system performance can be continuously optimized iteratively. However, we find that the retrained system still suffers from a variable number of occlusion and long-range perception errors. A potential explanation for this issue is the presence of a positioning error in the vehicle data from the V2V4Real dataset, as confirmed by its developer. This error may adversely affect the retraining process using data generated by \tool. Given the complex structure and specific communication mechanisms inherent in V2X cooperative perception systems, it is essential to integrate specific repair technologies to enhance the system's performance.

\subsection{Threats to Validity}

\textbf{Construct Validity.} First, \tool~exclusively conducts offline testing~\cite{DBLP:journals/ese/StoccoPT23} to identify performance issues related to occlusion and long-range detection in LiDAR-based cooperative object detection systems. Although offline testing is often indispensable for the development lifecycle of cooperative perception modules~\cite{haq2021can,gao2026fuzzing}, the present study does not assess the influence of cooperative perception module issues on subsequent tasks such as decision-making, planning, and control. Second, to evaluate the core capabilities of V2X cooperative perception systems in detecting occluded and long-range targets, V2XGen primarily focuses on occlusion and long-range perception errors, while not explicitly evaluating general scenes such as near-range or non-occluded cases. Although the current evaluation emphasizes these challenging conditions, the V2XGen framework is general and capable of generating diverse scenes. The framework can easily generate near-range, normal-range, and non-occluded interactive scenes by adjusting the fitness guidance. Furthermore, our technique focuses on LiDAR-based cooperative perception, and we leave multimodal fusion as future work due to the challenges it introduces, such as ensuring cross-modal semantic consistency.

\textbf{Internal Validity.} The primary internal threat arises from the data simulation process. Firstly, the parameter configuration of the virtual sensors in \tool~can significantly impact the quality of the synthesized data. Additionally, in the LiDAR simulation completion process of the deletion operator, this may introduce accuracy errors in meshifying the road point cloud to reconstruct the road surface. To mitigate these, we utilize real-world LiDAR configurations in V2X systems for our virtual sensor setup. Secondly, our design aims to ensure the diversity of inserted object shapes and enable the evaluation of the robustness of cooperative perception systems in long-tailed scenarios. However, the quality of the 3D models used in \tool~significantly affects the quality of the synthesized data. Since these data are generated by combining data from different datasets, the process may introduce additional distribution shifts. To mitigate this issue, we adopt ShapeNet, a widely used repository of 3D object assets, and employ a ray-casting algorithm to simulate the LiDAR emission process, thereby improving the realism of the synthesized data.

\textbf{External Validity.} First, the limitation of the experimental dataset poses a risk to the generalizability of the findings. The quality of the transformed data relies heavily on the characteristics of the original dataset. As a result, if the input data includes driving scenes not present in the training set, the retrained system may encounter difficulties in making accurate predictions. To address this concern, \tool~employs a variety of transformation operators and utilizes two widely used large-scale datasets for experiments, demonstrating the potential for generalization to other datasets. Second, variability in configurable transformation parameters poses an additional potential threat. Some of our findings may not apply to different parameter sets during evaluation. To mitigate this concern, we thoughtfully select rotation and scale parameters in our experimental design to closely replicate the actual cooperative driving environment. Regarding the translation parameter, due to the lack of fine-grained lane-level ground-truth information in the current scenario, the translation operator does not impose a specific parameter range but instead ensures that the transformed object remains on the road. Therefore, the direction of travel of the transformed object and the traffic flow may occasionally be inconsistent. However, our framework design is scalable if a high-resolution lane-level map is provided. Regarding the rotation parameter, we set the maximum rotation angle to 30$^\circ$ to cover challenging corner cases frequently encountered in real-world driving, such as skewed parking, aggressive lane-change maneuvers, or skidding, while acknowledging that this setting may introduce physically implausible scenarios. Third, given that the experimental dataset (e.g., V2V4Real) provides annotations only for vehicle-type entities, we restrict our evaluation to the most common entity type in V2X cooperative perception scenes. However, our operators (insertion, deletion, and composite transformations) rely on 3D LiDAR simulation and apply to other entity types. For the insertion operator, any 3D model (e.g., pedestrians or cyclists) can be added to the asset set $\mathbb{E}$ and rendered. For inserted entities, valid locations are determined via semantic segmentation (e.g., pedestrians and cyclists are placed on sidewalks, crosswalks, or shoulders based on corresponding labels). For the deletion operator, V2XGen performs background entity deletion based on the ground-truth label of the target entity and conducts LiDAR simulation completion using the LiDAR laser emission mechanism, ensuring independence from the specific entity type. The composite operator, which combines insertion and deletion operations, can also extend to other entity types.

\section{Related Work}

\textbf{Quality Assurance of Autonomous Driving Perception Systems.}
Perception systems are crucial for supporting ADS in industrial applications, significantly affecting overall system quality ~\cite{li2020lidar}. Research has focused on testing the robustness of single-agent perception systems across various driving tasks ~\cite{wang2020metamorphic,xie2022towards,DBLP:conf/issta/GuoF022,christian2023generating,gao2023benchmarking,DBLP:conf/icse/GaoW000X24,DBLP:conf/cvpr/ManivasagamWWZS20,DBLP:conf/iccv/ManivasagamBW0U23,DBLP:journals/tse/GuoSGFWTWMC26,DBLP:journals/pacmse/TianDHWGZCWZ25,10.1145/3832783.3837540}. To evaluate camera-based perception systems, a representative study~\cite{wang2020metamorphic} generates data by incorporating object instances into background images.
Several studies also test LiDAR-based perception systems. LiRTest ~\cite{DBLP:conf/issta/GuoF022} uses metamorphic relations to create test point clouds, identifying failures in 3D object detection models. Christian et al. ~\cite{christian2023generating} propose testing LiDAR systems by introducing semantic mutations to real-world data for the semantic segmentation task. Manivasagam et al.~\cite{DBLP:conf/cvpr/ManivasagamWWZS20,DBLP:conf/iccv/ManivasagamBW0U23} propose methods for simulating LiDAR point clouds by reconstructing digital twins of real-world scenes. However, these approaches rely on costly data collection, including extensive scanning of physical assets and HD maps, rather than performing entity-level transformations on existing real-world scenes. In contrast, V2XGen is a lightweight and automated framework that operates directly on existing point clouds and evaluates system correctness using predefined metamorphic relations.
Gao et al. present an object-insertion framework for multi-sensor fusion perception systems~\cite{DBLP:conf/icse/GaoW000X24}. It focuses solely on single-agent insertion, which is insufficient to generate diverse and challenging V2X scenarios. To address this limitation, V2XGen not only customizes insertion operators for V2X scenarios but also introduces a deletion operator and composite operators. Furthermore, extending these single-agent testing methods to cooperative perception systems becomes challenging due to the consistent nature of the multiple perspectives required for cooperative systems. Recently, Guo et al. first proposed CooTest~\cite{DBLP:conf/issta/GuoG00LGSF24}, a tool specifically designed to evaluate cooperative perception systems under extreme conditions (e.g., communication interference, adverse weather, etc.). CooTest lacks consideration of traffic participants and does not support flexible modifications to their number, styles, or spatial relationships within a scene. This limitation significantly reduces the diversity of the generated scenes. In contrast, \tool~addresses this gap by leveraging a series of operators to manipulate entities rigorously within scenes, enabling the flexible generation of realistic, diverse, and challenging traffic scenes that better reflect real-world V2X conditions. Furthermore, although CooTest provides valuable insights into system robustness, it does not explicitly evaluate functional correctness.

\noindent\textbf{Metamorphic Testing.}
Metamorphic Testing (MT)~\cite{segura2016survey,chen2018metamorphic} is a software testing methodology that alleviates the test oracle problem by identifying software bugs through the detection of deviations from domain-specific metamorphic relations defined across outputs from multiple executions of the program with various test inputs. 
With the rapid progression of deep learning, metamorphic testing has been applied in both classification and regression tasks~\cite{DBLP:conf/issta/GuoF022, xie2011testing, wang2020metamorphic, DBLP:conf/issta/GuoG00LGSF24,DBLP:journals/cacm/ZhouS19,DBLP:journals/jss/GuoFCC24,DBLP:journals/pacmse/ZhouCTGXXZ26}. Xie et al.~\cite{xie2011testing} introduced an MT technique to test machine learning classification algorithms, demonstrating its effectiveness in identifying errors within a widely recognized open-source classification program. In addition, Wang et al.~\cite{wang2020metamorphic} employed MT as an adaptive and effective testing strategy by comparing object detection results between original and synthetic images to uncover defects in object detection systems. Guo et al.~\cite{DBLP:conf/issta/GuoF022} introduced LiDAR-based specific metamorphic relations and utilized them to generate various corner test cases. Furthermore, CooTest~\cite{DBLP:conf/issta/GuoG00LGSF24} developed a specific V2X-oriented metamorphic testing module for cooperative perception systems.

\section{Conclusion}

In this paper, we present \tool, a tool for generating test scenes tailored specifically for V2X communication systems. \tool~utilizes a high-fidelity approach to render object instances and ground point clouds using a set of virtual sensors. It then synthesizes realistic cooperative perception data by simulating object instances in valid yet challenging locations within the target scene. Furthermore, \tool~incorporates fitness metric guidance to enhance testing efficiency and effectiveness. The experimental results on six state-of-the-art cooperative detection systems across various fusion schemes illustrate that \tool~adeptly detects erroneous behavior in the systems under test and enhances a system’s robustness through retraining.

\section*{Acknowledgments}
This work is supported partially by the National Key Research and Development Program of China (2024YFF0908000), Fundamental and Interdisciplinary Disciplines Breakthrough Plan of the Ministry of Education of China (No. JYB2025XDXM118), 111 Center (No. B26023), and the Open Project of State Key Laboratory for Novel Software Technology at Nanjing University under (Grant No. KFKT2024B21).


\bibliographystyle{ACM-Reference-Format-Order}
\bibliography{sample-base}










\end{document}